\documentclass[11pt,a4paper]{article}
\pdfoutput=1
\usepackage{jcappub}
\usepackage{ifpdf}
\usepackage{afterpage}
\usepackage{amssymb}
\usepackage{amsmath}
\usepackage{graphicx}
\usepackage{longtable}
\usepackage{verbatim}
\usepackage{amsfonts}
\usepackage{bbold}
\usepackage[table,usenames,dvipsnames]{xcolor}
\usepackage{titlesec}
\usepackage{cancel}
\usepackage{enumitem}
\usepackage{hyperref}
\usepackage{soul}
\usepackage{comment}
\usepackage{float}

\newcommand{\PI}{\mathcal{P}_\mathcal{II}^{(1)}}
\newcommand{\pii}{\mathcal{P}_\mathcal{II}^{(2)}}

\newcommand{\pr}{\mathcal{P}_\mathcal{RR}^{(1)}}
\newcommand{\prr}{\mathcal{P}_\mathcal{RR}^{(2)}}

\newcommand{\niso}{n_{\mathcal{II}}}
\newcommand{\fiso}{f_{\rm iso}}
\newcommand{\biso}{\beta_{\rm iso}}

\newcommand{\taureio}{\tau_{\rm reio}}
\newcommand{\logA}{\log (10^{10}A_{\rm s})}

\newcommand{\GeV}{\mathrm{GeV}}
\newcommand{\Mpc}{\mathrm{Mpc}}

\newcommand{\kc}{k_{\rm c}}
\newcommand{\kcm}{k_{\rm c,min}}
\newcommand{\ki}{k_{\rm init}}

\begin{document}

\begin{flushright}
CERN-TH-2026-136
\end{flushright}

\title{CMB Constraints on Pre-Inflationary Axion Dark Matter Isocurvature}
    
\author[1]{Catherine Petretti,}
\author[2]{Praniti Singh,}
\author[3]{Matteo Braglia,}     
\author[1]{Xingang Chen,}
\author[2,4]{JiJi Fan,}
\author[5,6]{and Lingfeng Li}
\affiliation[1]{Institute for Theory and Computation, Harvard-Smithsonian Center for Astrophysics, 60 Garden Street, Cambridge, MA 02138, USA}
\affiliation[2]{Department of Physics, Brown University, Providence, RI 02912, USA}
\affiliation[3]{Theoretical Physics Department, CERN, 1211 Geneva 23, Switzerland}
\affiliation[4]{Brown Center for Theoretical Physics and Science and Innovation, Brown University, Providence, RI 02912, USA}
\affiliation[5]{Nanjing Normal University, Nanjing, Jiangsu 210023, China}
\affiliation[6]{International Center for Theoretical Physics-Asia Pacific, University of Chinese Academy of Sciences, Beijing 100190, China}

\emailAdd{catherine.petretti@cfa.harvard.edu}
\emailAdd{praniti\_singh@brown.edu}
\emailAdd{matteo.braglia@cern.ch}
\emailAdd{xingang.chen@cfa.harvard.edu}
\emailAdd{jiji\_fan@brown.edu}
\emailAdd{l.f.li165@gmail.com}

\abstract
{Although measurements of the Cosmic Microwave Background (CMB) are consistent with a nearly scale-invariant primordial spectrum of adiabatic perturbations, in which the energy densities of different components (radiation, baryons, and dark matter) fluctuate proportionally, there could also exist isocurvature perturbations, in which density fluctuations of the individual components differ from the adiabatic mode. Cold dark matter isocurvature (CDI) perturbations with a variety of spectral tilts generated in pre-inflationary axion models provide one such example. In this article, we present the most updated constraints on these axion CDI perturbations using the latest CMB anisotropy measurements from Planck, the Atacama Cosmology Telescope (ACT), and the South Pole Telescope (SPT). We study both fixed spectral indices with values ranging from red- to blue-tilted spectra as well as the case with a free index. We find that the constraint on the spectral index gets moderately improved with the combined datasets compared to Planck alone, while the bounds on the isocurvature amplitudes for the fixed spectral indices we consider do not get tighter. We also discuss the theoretical implications of our constraints, in particular for models giving rise to blue-tilted spectra.}
\maketitle
    
\section{Introduction}
\label{sec:intro}

Measurements of the Cosmic Microwave Background (CMB) over decades lay the solid foundation for modern cosmology. One of the key findings from CMB observations~\cite{Planck:2018vyg, Planck:2018jri} is an almost scale-invariant primordial spectrum of the curvature perturbation mode (also known as the adiabatic mode), in which the energy densities of different components such as radiation and matter fluctuate proportionally. In addition to the curvature/adiabatic mode, there could potentially exist non-adiabatic perturbations, the so-called isocurvature modes. These perturbations of different fluids, including cold dark matter (DM), baryonic matter, and radiation, deviate from the adiabatic one. For example, if the cold DM fluctuation is different from the curvature one, the corresponding isocurvature mode is called cold DM isocurvature (CDI). If all the radiation and matter contents are generated from the reheating of a single inflaton field, the isocurvature mode would be highly suppressed in CMB observations. On the other hand, if multiple independent perturbations are present during inflation and pass onto different components of our observable universe, isocurvature modes could be non-negligible and serve as powerful probes of new physics beyond the minimal single-field inflation scenario and the Standard Model of particle physics. 

Recently, data releases from the Atacama Cosmology Telescope (ACT)~\cite{ACT:2025tim} and the South Pole Telescope (SPT)~\cite{SPT-3G:2025bzu} allow us to further probe and constrain CDI with the CMB. In particular, given their significantly improved sensitivities to the CMB at large angular multipole moments $\ell$'s (smaller length scales), ACT and SPT could potentially provide us with more information about the CDI spectrum with a spectral index above one, the ``blue-tilted" spectrum with more power at larger $\ell$'s.
 
One highly-motivated new physics scenario with a potentially large CDI is the pre-inflationary axion DM model. Axions can arise from various types of field theory models~\cite{Kim:1979if, Shifman:1979if,Zhitnitsky:1980tq, Dine:1981rt} or compactifications of higher-dimensional, higher-form gauge fields in string theory~\cite{Witten:1984dg}. In addition, one particular type of axion, the QCD axion, is a leading solution to the strong CP problem in the Standard Model of particle physics~\cite{Peccei:1977hh, Peccei:1977ur,Weinberg:1977ma,Wilczek:1977pj}. The axion also serves as a classic benchmark model of cold DM~\cite{Preskill:1982cy,Dine:1982ah,Abbott:1982af}. In the early universe, once the Hubble expansion rate drops to around the axion mass, the axion, initially at rest in its field space due to Hubble friction, could oscillate coherently around the minimum of its potential. Its coherent oscillation redshifts as non-relativistic matter and contributes to the abundance of cold DM. Given their broad range of applications in cosmology and particle physics, axions have long been the subject of considerable interest.

The cosmological history of axion DM falls into two categories: pre-inflationary and post-inflationary ones. Roughly speaking, in the first class, the massless axion already exists during inflation, while in the latter case, the axion is not present during inflation (for a review, see~\cite{Marsh:2015xka}). In this article, we will focus on the pre-inflationary axion DM scenario. This is motivated theoretically, in particular for QCD axion DM, as the pre-inflationary scenario could be more compatible with solutions to the QCD axion quality problem compared to the post-inflationary scenario~\cite{Demirtas:2021gsq,Lu:2023ayc,Benabou:2023npn, Reece:2024wrn}. Since the pre-inflationary axion is already generated during inflation, it fluctuates independently of the inflaton fluctuations. These primordial fluctuations of the axion field would turn into CDI once the axion becomes cold DM after inflation through the misalignment mechanism. 

So far, no isocurvature mode has been observed in the CMB data. Instead, there are strong limits on different isocurvature modes, including constraints on CDI from the Planck satellite. For a scale-invariant primordial CDI spectrum, its amplitude is bounded to be at most $8 \times 10^{-11}$, about $4\%$ of the curvature mode's amplitude, at 95\% confidence level (C.L.)~\cite{Planck:2018jri}.\footnote{Constraints on other types of isocurvature spectra could be found in~\cite{Buckley:2025zgh}.} This tight constraint could be translated into very informative restrictions on pre-inflationary axion DM models. In particular, it leads to the (in)famous isocurvature problem of the QCD axion: if the QCD axion is the dominant component of DM, it is incompatible with high-scale inflationary scenarios with a Hubble scale above $10^7~\GeV$, assuming an ${\cal O}(1)$ initial misalignment angle of the axion field and the vanilla misalignment mechanism.

In this article, we aim to set the most updated constraints on pre-inflationary axion CDI in cases with either a flat or a tilted primordial spectrum with a fixed spectral index~\cite{Kasuya:1996ns,Kasuya:2009up,Chung:2016wvv,Chung:2021lfg,Chung:2023xcv,Chung:2024ctx,Bodas:2025eca,Redi:2022llj,Kalia:2025uxg}, using the latest data releases of ACT and SPT, combined with the Planck dataset. We will also study the case allowing both the spectral index and isocurvature amplitude to vary. In addition, the theoretical implications for models leading to blue-tilted spectra will be examined. 

The paper is organized as follows: in Sec.~\ref{sec:axion}, we review the basic formalisms of pre-inflationary axion DM and its corresponding CDI. In Sec.~\ref{sec:data}, we present the CMB data and methods used in the analysis. In Sec.~\ref{sec:results}, we present the results as constraints on pre-inflationary axion CDI and discuss possible improvements with future data. In Sec.~\ref{sec:axionconstraints}, we study the implications of these constraints for different axion models. We conclude in Sec.~\ref{sec:con}.

\section{Review of Axion CDI}
\label{sec:axion}

In this section, we will provide a brief review of the axion CDI basis, using the QCD axion as our benchmark. The discussions here and results presented later could be easily generalized to axion-like particles. The axion is denoted as $a$, a periodic CP-odd real scalar. For convenience, we define a dimensionless quantity during inflation $\theta=a/f_I$. Note that $f_I$, the axion decay constant during inflation, could be different from the axion decay constant after inflation, denoted as $f_a$. In the pre-inflationary axion scenario, an axion is present during inflation and fluctuates. The amplitude of the axion fluctuation is
\begin{equation}
   \delta \theta = \theta - \langle \theta \rangle = \theta-\theta_i~, 
\end{equation}
where $\langle\theta \rangle = \theta_i$ is the initial misalignment angle, which is set by the Gibbons-Hawking effect~\cite{Gibbons:1977mu}:
\begin{equation}
    \sigma_\theta = \sqrt{(\delta \theta)^2} = \frac{H_I}{2\pi f_I}~,
\end{equation}
where $H_I$ is the Hubble scale during inflation.

Around the QCD phase transition after inflation, the axion starts to gain a temperature-dependent potential, $V_a(T)$, from non-perturbative QCD dynamics. Once the Hubble rate $H(T)$ drops to around the axion mass $m_a(T)$, the axion field starts to oscillate and contributes to the DM relic density through the misalignment mechanism~\cite{Preskill:1982cy,Dine:1982ah,Abbott:1982af}. The primordial axion fluctuation, independent of the inflaton one, is then converted into CDI. The isocurvature fluctuation in DM, $S_d$, and the contribution from the axion component, $S_a$, are defined as
\begin{equation}
    S_d \equiv \frac{\delta \Omega_d}{\Omega_d}-\frac{3}{4}\frac{\delta \Omega_{\rm rad}}{\Omega_{\rm rad}}~,\quad S_a \equiv \frac{\delta \Omega_a}{\Omega_a}-\frac{3}{4}\frac{\delta \Omega_{\rm rad}}{\Omega_{\rm rad}}~,
\end{equation} 
where $\delta \Omega$ indicates the fluctuation of each component's abundance, and $\Omega$ is the average value. The subscripts $d$, $a$, and rad refer to DM, axion, and photon radiation, respectively.\footnote{A most recent analytical explanation of axion perturbations at cosmological scales applicable to a wide range of cosmological axion scenarios can be found in~\cite{Allali:2025pja}. } 
In this work, we assume that axion DM isocurvature is the only contribution to CDI. The fluctuations in the rest of DM have the same origin as those of radiation and do not contribute to CDI. Thus
\begin{equation}
    S_d = \gamma S_a~,
\end{equation}
where $\gamma \equiv \Omega_a/\Omega_d$ is the fraction of axion DM relic abundance $\Omega_a h$ relative to the total DM abundance $\Omega_d h$.

The dimensionless primordial CDI spectrum is defined by the two-point correlator of the CDI mode $S_d$ as
\begin{equation}
    \frac{k^3}{2\pi^2} \langle  S_d(\textbf{k}) S_d(\textbf{k}^\prime) \rangle \equiv (2\pi)^3 \delta^{(3)}(\textbf{k}-\textbf{k}^\prime) \mathcal{P}_\mathcal{II}(k)~,
\end{equation}
with the following power-law parameterization:
\begin{equation}
    \mathcal{P}_\mathcal{II}(k) = A_\mathcal{I} \left(\frac{k}{k_*}\right)^{\niso-1}~,
    \label{eq:nidefinition}
\end{equation}
where $A_\mathcal{I}$ is the amplitude, $k_*$ is an arbitrary pivot scale (often taken to be $k_*=0.05~\Mpc^{-1}$), and $\niso$ is the CDI spectral index. As we will discuss more in Sec.~\ref{sec:data}, this parametrization, which is widely used in theoretical axion studies, is not the one usually used in CMB data analyses. Yet, it is straightforward to convert results between different parameterizations. It is also useful to define the isocurvature fraction
\begin{equation}
    \label{eq:beta_iso}
    \biso(k)\equiv\frac{\fiso^2(k)}{1+\fiso^2(k)}~,
\end{equation}
where $\fiso^2(k) \equiv \frac{\mathcal{P}_\mathcal{II}(k)}{\mathcal{P}_\mathcal{RR}(k)}$ is the ratio between the power spectra of isocurvature perturbations $\mathcal{P}_\mathcal{II}$ and adiabatic perturbations $\mathcal{P}_\mathcal{RR}$ at a given scale $k$. For small $\fiso^2$, we have $\biso\simeq \fiso^2$.

If the primordial axion spectrum is almost flat (nearly scale-invariant) with $\niso\approx 1$, 
then the CDI amplitude $A_\mathcal{I}$ relevant for CMB analysis is given by~\cite{Kobayashi:2013nva, Allali:2025pja}
\begin{equation}
    A_{\mathcal{I}} =  \left(\gamma\sigma_{\theta} \frac{\partial \ln \Omega_a}{\partial  \theta_i}\right)^2=\left(\frac{\gamma H_I}{2\pi f_I } \frac{\partial \ln \Omega_a}{\partial  \theta_i}\right)^2~,
    \label{eq:Ai}
\end{equation}
in which $\partial \ln \Omega_a/\partial \theta_i$ takes into account the anharmonic effects due to the full axion potential beyond the quadratic approximation, which is particularly important in the large misalignment scenario when the axion starts to oscillate near the hilltop of its potential. The axion relic abundance $\Omega_a$ is given by~\cite{Lyth:1991ub,Strobl:1994wk,Bae:2008ue, Visinelli:2009zm,Dine:2017swf}
\begin{eqnarray}
    \Omega_a &\simeq& 0.27\times \tilde{\theta}_i^2 \;\bigg(\frac{f_a}{10^{12}~\GeV}\bigg)^\frac{7}{6}~, \nonumber \\
    \tilde{\theta}_i^2 &= & \theta_i^2 \; \log\bigg[\frac{e}{1-(\theta_i^2/\pi^2)}\bigg]^{\frac{7}{6}}~,
    \label{eq:abundance}
\end{eqnarray}
where $\tilde{\theta}_i$ accounts for the anharmonic effects which become important when $|\theta_i| \sim \pi$. 

The primordial axion spectrum could also be either ``red" ($\niso<1 $) or ``blue" ($\niso >$1) tilted, depending on the dynamics of the PQ field during inflation. One well-quoted example is the supersymmetric PQ model with a Hubble induced mass term~\cite{Kasuya:2009up}. In this type of model, the value of the PQ field decreases during inflation, resulting in a blue-tilted isocurvature spectrum with $2 \lesssim \niso \lesssim 4$. Some further studies of this model and other model building attempts to have $\niso> 1$ can be found in~\cite{Kasuya:1996ns, Chung:2016wvv,Chung:2021lfg,Chung:2023xcv,Chung:2024ctx,Bodas:2025eca,Redi:2022llj,Kalia:2025uxg}. Note that these studies are not limited to axion DM and their parameterizations vary.

\section{Data and Methods}
\label{sec:data}

\subsection{Primordial Spectra Parametrization}
\label{subsec:notation}
A general combination of  primordial curvature (adiabatic) and isocurvature fluctuations is characterized by the curvature power spectrum $\mathcal{P}_\mathcal{RR}(k)$, the isocurvature power spectrum $\mathcal{P}_\mathcal{II}(k)$, and their cross-spectrum $\mathcal{P}_\mathcal{RI}(k)$. Following the analyses of Planck~\cite{Planck:2018jri} and ACT~\cite{ACT:2025tim}, we adopt a power law for the primordial spectra, parameterized by their amplitudes at two reference scales, $k_1=0.002~\Mpc^{-1}$ and $k_2=0.1~\Mpc^{-1}$. The spectra at arbitrary scales are given by: 

\begin{equation}
    {\cal P}_{XY}(k)=\exp
    \Biggl[
    \left(
    \frac
    {\ln (k  )-\ln (k_2)}
    {\ln (k_1)-\ln (k_2)}
    \right)
    \ln( {\cal P}^{(1)}_{XY})
    +
    \left(
    \frac
    {\ln (k  )-\ln (k_1)}
    {\ln (k_2)-\ln (k_1)}
    \right)
    \ln( {\cal P}^{(2)}_{XY})
    \Biggr],
    \label{eq:two-scale}
\end{equation}
where $\mathcal{P}_{XY}^{(i)} \equiv \mathcal{P}_{XY}\left(k_i\right) \text { and } X, Y \in\{\mathcal{R}, \mathcal{I}\}$. We further emphasize that, in our analysis, the isocurvature fluctuations arise exclusively from CDI modes and are assumed to be completely uncorrelated with the adiabatic mode such that $\mathcal{P}_{\mathcal{RI}}^{(i)}=0$.  

The two-scale amplitude parameterization of the primordial isocurvature spectrum has a one-to-one mapping with the simple power-law parameterization Eq.~\eqref{eq:nidefinition} in axion studies. For the explicit mapping between these parameterizations, see App.~\ref{app:notations}. 

\begin{figure}[ht]
    \centering
    \includegraphics[width=\linewidth]{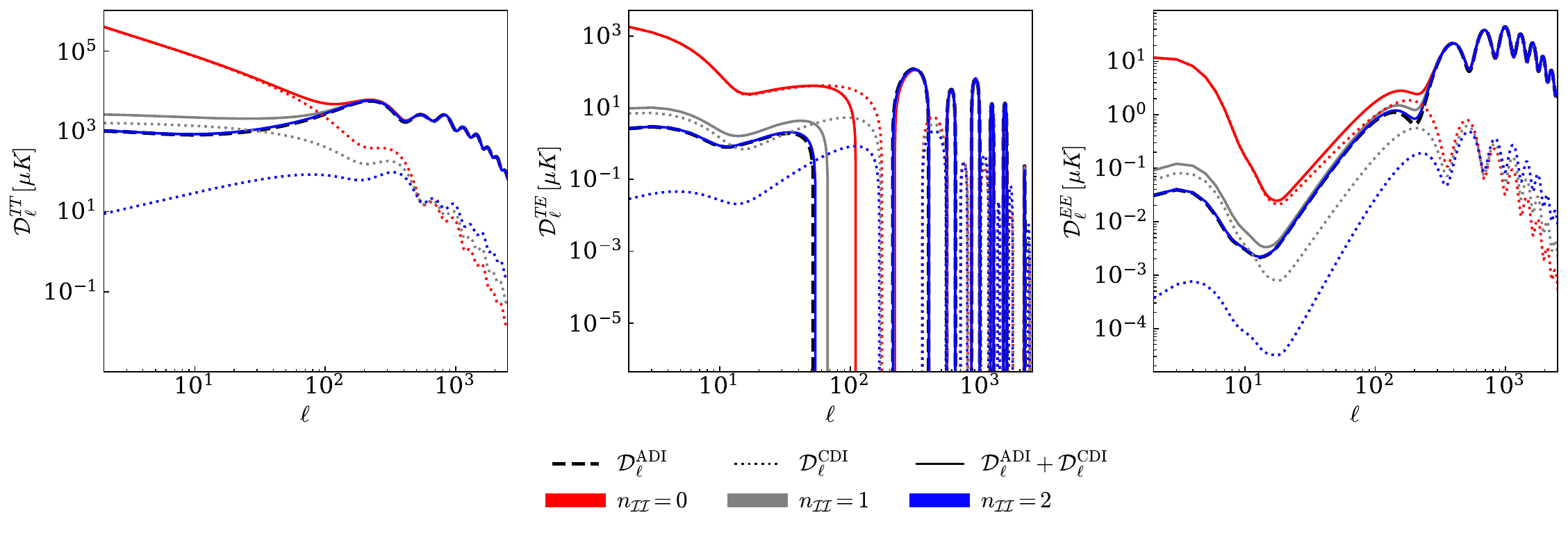}
    
    \caption{\textit{TT}, $EE$, and \textit{TE} angular spectra of CMB anisotropies for uncorrelated adiabatic (denoted as ADI in the plot) and CDI initial conditions. The dashed line shows the adiabatic-only spectra with $\logA=3.045$ and $n_{\rm s}=0.965$, and the dotted lines show CDI-only spectra with three different $\niso$'s. For comparison, we choose $A_\mathcal{I}=A_{\rm s}$. The solid lines show the sum of the CDI and adiabatic spectra. Note that the dashed line representing ADI almost overlaps with the sum of ADI and CDI with $\niso=2$. }
    \label{fig:power_spectrum}
\end{figure}

In Figure~\ref{fig:power_spectrum}, we plot the angular power spectra of the CMB anisotropies induced by a mixture of uncorrelated adiabatic and CDI initial conditions for a few values of $\niso$ and choosing the CDI spectrum amplitude, $A_\mathcal{I}$, to be the same as the curvature spectrum amplitude, $A_s$. As can be seen, the CDI contribution to the spectra decreases at high multipoles due to the shape of the CDI transfer functions. CDI mostly contributes at large scales, unless the spectral index $\niso\gg 1$.

\subsection{Data Analysis} \label{subsec:data}
The benchmark scenarios we consider either have fixed CDI spectral indices $\niso=\{0,1,2,3 \}$ or a free $\niso$.\footnote{For axion models discussed in Sec.~\ref{sec:axion} and Sec.~\ref{sec:axionconstraints}, $\niso$ benchmarks with integer values other than 1 are not special. However, for other models where the CDI are generated by small-scale stochastic behavior, at large scales the isocurvature perturbation's spatial correlation will be scale-independent, which corresponds to the extremely blue-tilted isocurvature spectrum with $\niso =4$.} In the two-scale parameterization for a fixed $\niso$, $\PI$ is a free parameter and $\pii$ is derived according to the value of $\niso$, while for free $\niso$, both $\PI$ and $\pii$ are treated as free parameters.

Besides the power spectrum parameters $\{\pr,~\prr,~\PI,~\pii \}$, other free cosmological parameters in our analyses include: the Hubble rate of expansion today (Hubble constant) $H_0=100 \, h$~km/s/Mpc, the baryon density $\omega_b=\Omega_b h^2$ (where $\Omega_i\equiv \rho_i/\rho_{c}$ is the abundance for the $i^\text{th}$ species, $\rho_i$ its density, and $\rho_{c}$ the critical energy density for a flat universe today), the cold DM density $\omega_c=\Omega_c h^2$, and the reionization optical depth $\taureio$. The priors placed on the cosmological and primordial spectrum parameters are listed in Table~\ref{tab:priors}. Additionally, following Planck's conventions~\cite{Planck:2018vyg}, we model the Standard Model neutrinos as a combination of one massive species with a mass of 0.06 eV and two massless species, enforcing $N_{\rm eff}=3.046$.

\begin{table}[ht]
    \centering
    \begin{tabular}{|l|l|}
         \hline
         Parameters & Priors \\ \hline
         \multicolumn{2}{|c|}{Cosmological Parameters} \\ \hline
        
        $h$ & $[0.6,0.8]$ \\ \hline
        $\Omega_bh^2$ & $[0.02,0.0265]$ \\ \hline
        $\Omega_c h^2$ & $[0.1,0.135]$ \\ \hline
        $\taureio$ & $[0.01,0.8]$ \\ \hline
        \multicolumn{2}{|c|}{Adiabatic and CDI Parameters} \\ \hline
        $\pr$ & $[1.5,4.0]\times 10^{-9}$ \\ \hline
        $\prr$ & $[1.5,4.0]\times 10^{-9}$ \\ \hline
        $\PI$ & $[0,1.0]\times 10^{-8}$ \\ \hline
        $\pii$ (for $\niso=$ free)& $[0,1.0]\times 10^{-8}$ \\ \hline
    \end{tabular}
    \caption{Priors placed on the free parameters in our analysis. We place uniform priors on the four cosmological parameters unrelated to the primordial power spectrum ($h$, $\Omega_bh^2$, $\Omega_c h^2$, $\taureio$) as well as the adiabatic and isocurvature spectrum parameters. Here, we report the upper and lower bounds of the priors.}
    \label{tab:priors}
\end{table}

We employ the Boltzmann solver code {\tt CLASS}~\cite{Diego_Blas_2011,lesgourgues2011:cos} to solve for the cosmic evolution of the parameter space and use {\tt Cobaya}~\cite{Torrado:2020dgo} to perform nested sampling via {\tt PolyChord}~\cite{Handley:2015fda,Handley_2015}. Posteriors are obtained using {\tt GetDist}~\cite{lewis2019getdistpythonpackageanalysing}. We use {\tt HyRec}~\cite{Ali-Haimoud:2010hou} to compute recombination physics and {\tt HMcode}~\cite{Mead:2020vgs} to calculate the non-linear corrections to the matter power spectrum.\footnote{We also set the parameter {\tt hmcode\_max\_k\_extra=100} in {\tt CLASS}, meaning that the matter power spectrum will be computed for 100 additional $k$-values beyond the maximum $k$-value. This setting is crucial for ensuring accuracy when computing blue-tilted isocurvature spectra.}

In order to study CDI fluctuations and constrain $\mathcal{P}_\mathcal{II}(k)$, we use CMB observations from Planck, ACT Data Release 6 (DR6), and SPT-3G D1. We use the following likelihoods: 
\begin{itemize}
    \item For low-$\ell$ Planck \textit{TT} data, we use the {\tt Commander}-based likelihood from the Planck Public Release 3 (PR3), which covers multipoles $\ell<30$~\cite{Planck:2019nip}.
    
    \item For low-$\ell$ Planck \textit{EE} data,  we use the {\tt Sroll}2 likelihood,\footnote{\url{https://web.fe.infn.it/~pagano/low_ell_datasets/sroll2/}} also covering multipoles $\ell<30$~\cite{Pagano_2020}.
    
    \item For high-$\ell$ Planck data, we use the Planck PR4 {\tt NPIPE}~\cite{Planck:2020olo} version of the {\tt CamSpec} likelihood~\cite{Efstathiou:2019mdh}, which  covers the multipole ranges $\ell \in [30,2500]$ for \textit{TT} and $\ell \in [30,2000]$ for \textit{TE}/\textit{EE}~\cite{Rosenberg}.
    
    \item For ACT DR6 \textit{TT}, \textit{TE}, and \textit{EE} spectra, we use the multi-frequency {\tt MFLike}~likelihood, covering the range $\ell \in [600,8500]$~\cite{ACT:2025fju}. 

    \item For SPT-3G D1 \textit{TT}, \textit{TE}, and \textit{EE} spectra, we use the {\tt candl} likelihood~\cite{SPT-3G:2025bzu}, covering $\ell \in [400 ,3000]$ for \textit{TT} and $\ell \in [400 ,4000]$ for \textit{TE}/\textit{EE}.
\end{itemize}
We combine these likelihoods into the following three datasets for our analyses:

\begin{itemize}
    \item \textbf{Planck}: contains only CMB observations from the Planck satellite, using the \textit{TT} {\tt Commander}, \textit{EE} {\tt Sroll}2, and {\tt NPIPE~CamSpec} likelihoods.

    \item \textbf{P-ACT}: a combination of Planck and ACT data. Here, we use the full Planck \textit{TT} {\tt Commander}, \textit{EE} {\tt Sroll}2, and ACT {\tt MFLike} likelihoods, and following~\cite{ACT:2025fju}, we cut the {\tt CamSpec} likelihood to only include multipoles $\ell < 1000$ for \textit{TT} and $\ell < 600$ for \textit{TE}/\textit{EE} data. These cuts are applied in order to reduce the sky overlap between the two datasets in the multipole ranges where they both have data. We choose ACT over Planck in these multipole ranges since ACT has similar precision in the overlap regions compared to Planck in \textit{TT} and higher precision in \textit{EE}. 

    \item \textbf{P-SPT}: a combination of Planck and SPT data. Similar to P-ACT, we use the full Planck low-$\ell$ \textit{TT} and low-$\ell$ \textit{EE} likelihoods, and we restrict
    the Planck high-$\ell$ data to $\ell < 2000$ for \textit{TT} and $\ell < 1200$ for \textit{TE}/\textit{EE}. We also restrict the SPT-3G D1 {\tt candl} likelihood to $\ell \geq 1200$ for both temperature and polarization spectra. We choose these cuts to favor the higher signal-to-noise in Planck or SPT at different scales. Although there is some overlap between the SPT and Planck data in the \textit{TT} channel for our choices of $\ell$, since the sky fraction of SPT is small, we can approximate these observations to be uncorrelated, as done in~\cite{SPT-3G:2025bzu}.
\end{itemize}
We highlight that while our analyses employs the {\tt CamSpec} likelihood for high-$\ell$ Planck data, the ACT collaboration uses the Planck {\tt plik\_lite} likelihood~\cite{Planck:2019nip}, which is marginalized over nuisance parameters~\cite{ACT:2025fju,ACT:2025tim}. Due to this difference, we anticipate that our constraints on CDI parameters obtained using the Planck likelihood are tighter than those reported by the ACT collaboration. We include a more thorough discussion and various tests on this likelihood difference in App.~\ref{app:likelihoods}.

To understand the impacts of ACT or SPT data in the combined datasets, we also have two separate ACT and SPT analyses for free $\niso$. For the \textbf{ACT} dataset, we use the ACT DR6 {\tt MFLike} likelihood as well as the Planck {\tt Sroll}2 likelihood to provide a Planck-derived measurement of the reionization optical depth $\taureio$ following~\cite{ACT:2025tim}. Similarly for the \textbf{SPT} dataset, we use the foreground-marginalized CMB-only SPT-3G D1 {\tt candl-lite} version~\cite{SPT-3G:2025bzu} along with the Planck {\tt Sroll}2 likelihood.

We note that CDI are also constrained by COBE/FIRAS limits on CMB spectral distortions~\cite{Mather:1993ij,Fixsen:1996nj}, as the dissipation of small-scale isocurvature fluctuations drives the CMB photons away from the thermal distribution, creating both $\mu$- and $y$-type distortions. In particular, since spectral distortions are sensitive to smaller scales than CMB anisotropies, the signal is enhanced for highly blue-tilted CDI, leading to strong constraints~\cite{Chluba:2013dna}.\footnote{The spectral distortion constraints on CDI cannot apply directly when Eq.~\eqref{eq:nidefinition} has a cutoff scale $k_{\rm cut}$ above which $\mathcal{P}_\mathcal{II}$ ceases to grow following the power law. This is the case for many realistic scenarios, see e.g.~\cite{Chung:2021lfg}. For example, the $\mu$-type distortion is most sensitive to CDI modes at a comoving scale $k\sim \mathcal{O}(10)~\Mpc^{-1}$, depending on the $\niso$ value. A cutoff scale around or smaller than this further weakens the constraints, especially those from the $\mu$-distortion whose sensitivity extends to smaller scales than the $y$-type one.} However, we have found that the constraints from COBE/FIRAS on the models studied in this paper---both the benchmarks with fixed $0\leq \niso\leq 3$ as well as with free $\niso$---are not competitive compared to those from CMB anisotropies. For more details, see App.~\ref{app:SDs}.

\section{Results}
\label{sec:results}

\begin{table}[ht]
    \small
    \centering
    \begin{tabular}{|l|llllll|}
        \hline Model \& Data &  $\PI$ & $ \pii$ & $ 100~\biso^{(1)}$ & $ 100~\biso^*$ & $ 100~\biso^{(2)}$ & $\niso$ \\
        
        \hline \multicolumn{7}{|l|}{ $\niso=0$} \\
        \hline 
        Planck & $< 5.07\times 10^{-11}      $ & $< 1.01\times 10^{-12}      $ & $< 2.11$ &  $< 0.0960$ & $< 0.0492$ & $\hdots$\\
        P-ACT & $< 4.63\times 10^{-11}$ & $< 9.27\times 10^{-13}$ & $< 1.94$ & $< 0.0861$ & $< 0.0439$ &  $\hdots$\\
        P-SPT & $< 4.51\times 10^{-11}$ & $< 9.02\times 10^{-13}$ & $< 1.90$ & $< 0.0854$ & $< 0.0435$ &  $\hdots$ \\
        
        \hline \multicolumn{7}{|l|}{ $\niso=1$} \\
        \hline 
        Planck & $< 5.30\times 10^{-11}$ & $< 5.30 \times 10^{-11}$ & $< 2.21$ & $< 2.45$ & $< 2.51$ & $\hdots$\\
        P-ACT & $< 7.51 \times 10^{-11}$ & $< 7.51 \times 10^{-11}$ & $< 3.12$ & $< 3.36$ & $< 3.41$ & $\hdots$\\
        P-SPT & $< 5.72\times 10^{-11}$ & $< 5.72 \times 10^{-11}$ & $< 2.39$ & $< 2.62$ & $< 2.67$ & $\hdots$\\
        
        \hline \multicolumn{7}{|l|}{ $\niso=2$} \\
        \hline 
        Planck &  $< 1.18\times 10^{-11}$ & $< 5.88\times 10^{-10}$ & $< 0.500     $ & $<  12.2 $  & $<22.2$ & $\hdots$\\
        P-ACT & $< 1.68\times 10^{-11}$ & $< 8.38\times 10^{-10}$ & $< 0.716$ & $< 16.3$ & $< 28.3$ & $\hdots$\\
        P-SPT & $< 1.27\times 10^{-11}$ & $< 6.34\times 10^{-10}$ & $< 0.540     $ & $<  13.0 $  & $<23.3$ & $\hdots$\\

        \hline \multicolumn{7}{|l|}{ $\niso=3$} \\
        \hline 
        Planck & $< 1.14\times 10^{-12}$ & $< 2.86\times 10^{-9}$ & $< 0.0491$ & $< 25.4$ & $< 58.3$ & $\hdots$\\
        P-ACT & $< 1.33\times 10^{-12}$ & $<3.34\times 10^{-9}$ & $< 0.0575$ & $< 28.0$ & $< 61.3$ & $\hdots$ \\
         P-SPT & $< 1.28\times 10^{-12}$ & $< 3.19\times 10^{-9}$ & $< 0.0549$ & $< 27.4$ & $< 60.6$ & $\hdots$\\
        \hline \multicolumn{7}{|l|}{$\niso$ free} \\\hline 
        Planck &  $ < 3.63 \times 10^{-11}$ & $< 2.84 \times 10^{-9} $ & $< 1.54$ & $< 23.9$  & $< 58.2$ & $2.2^{+1.4}_{-1.4}$ \\
        P-ACT  & $<5.01 \times 10^{-11}$ & $<2.79 \times 10^{-9}$ & $<  2.09$ & $< 25.1$ & $<56.8$ & $2.0^{+1.3}_{-1.1}$ \\
        P-SPT & $< 4.19\times 10^{-11}      $& $< 2.90\times 10^{-9}       $& $<1.78$ & $<25.5$ & $<58.2$&$2.1^{+1.2}_{-1.2}         $ \\
        \hline
    \end{tabular}
        \caption{95\% C.L. constraints on CDI model parameters from the Planck,~P-ACT,~and P-SPT datasets. $\biso^{(1)} = \biso(k_1)$, $\biso^{(2)} = \biso(k_2)$, and $\biso^{*} = \biso(k_*)$ at three different reference scales introduced in the main text $k_1=0.002$ Mpc$^{-1}$, $k_2 = 0.1$ Mpc$^{-1}$ and $k_*=0.05$ Mpc$^{-1}$.}
    \label{tab:constraints}
\end{table}

The results from the Planck, P-ACT, and P-SPT datasets are summarized in Table~\ref{tab:constraints}. We report the constraints on the primary isocurvature parameters, $\PI$ and $\pii$, as well as derived constraints on the primordial isocurvature fraction $\biso(k)$ (see Eq.~\eqref{eq:beta_iso}) at three reference scales, $\{k_1, k_*, k_2\}$, where $k_* = 0.05~\Mpc^{-1}$. For completeness, we also report constraints on all of the cosmological parameters in App.~\ref{app:full_constraints}. 

\subsection{Results for Spectra with a Fixed Tilt}

\begin{figure}[ht]
    \centering
    \includegraphics[width=0.49\columnwidth]{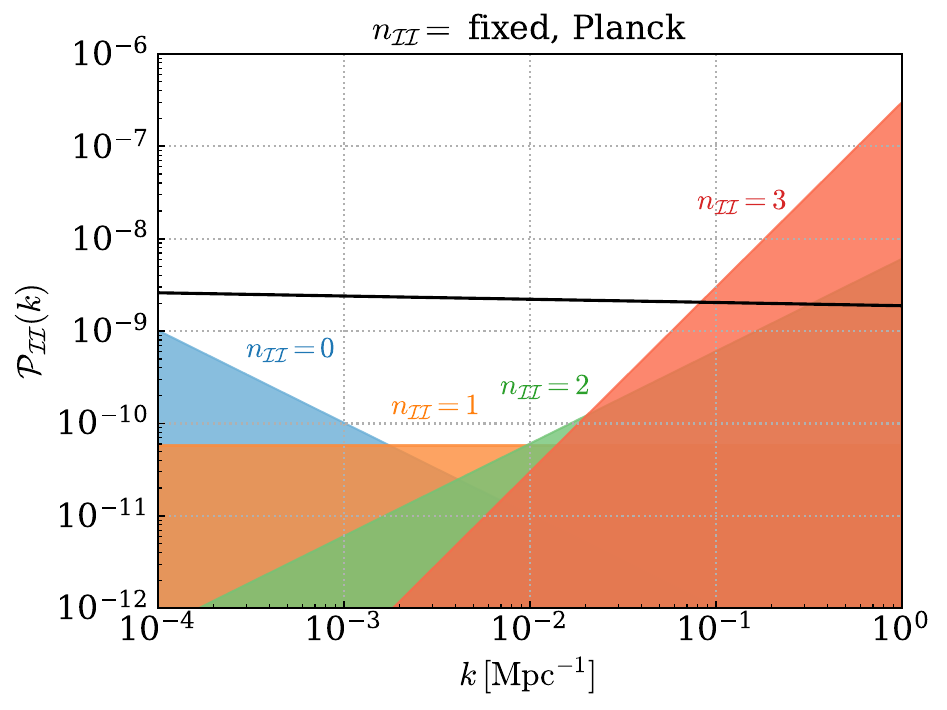}
    \hfill
    \includegraphics[width=0.49\columnwidth]{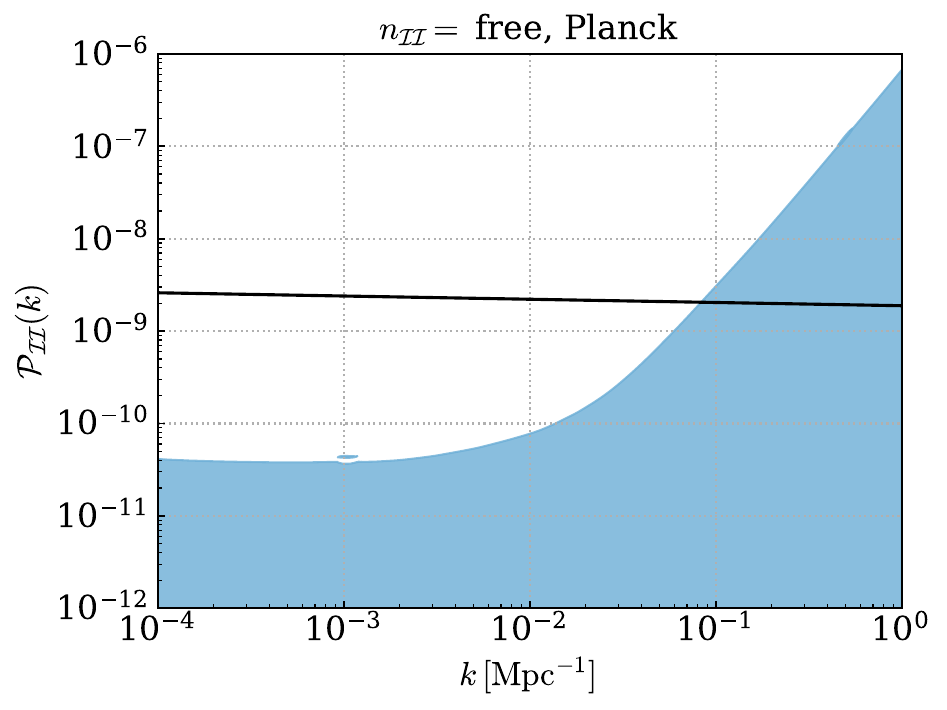}
    \includegraphics[width=0.49\columnwidth]{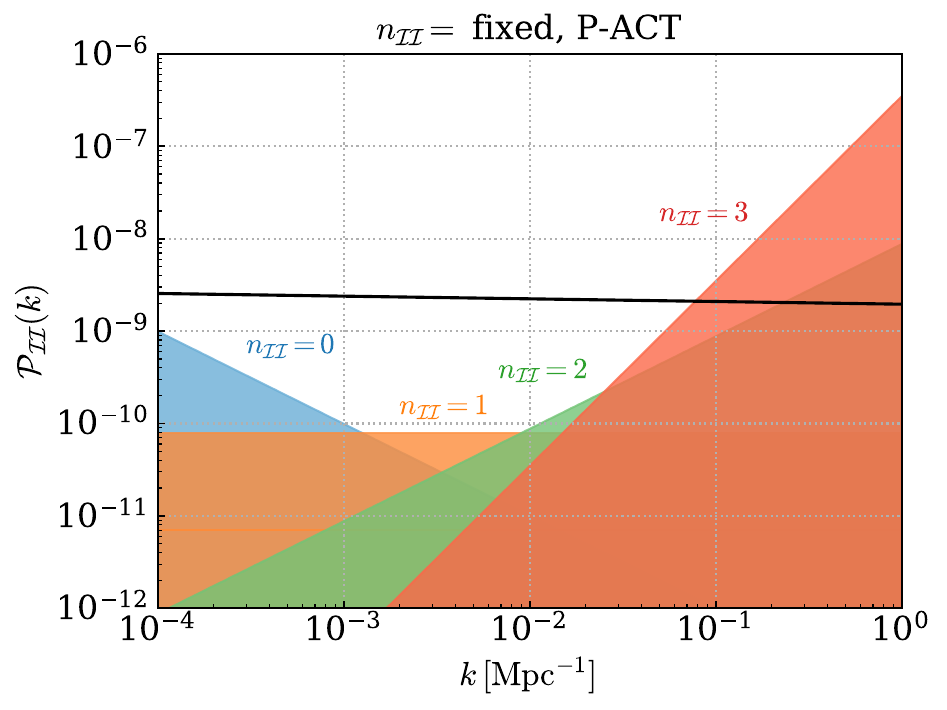}
    \hfill
    \includegraphics[width=0.49\columnwidth]{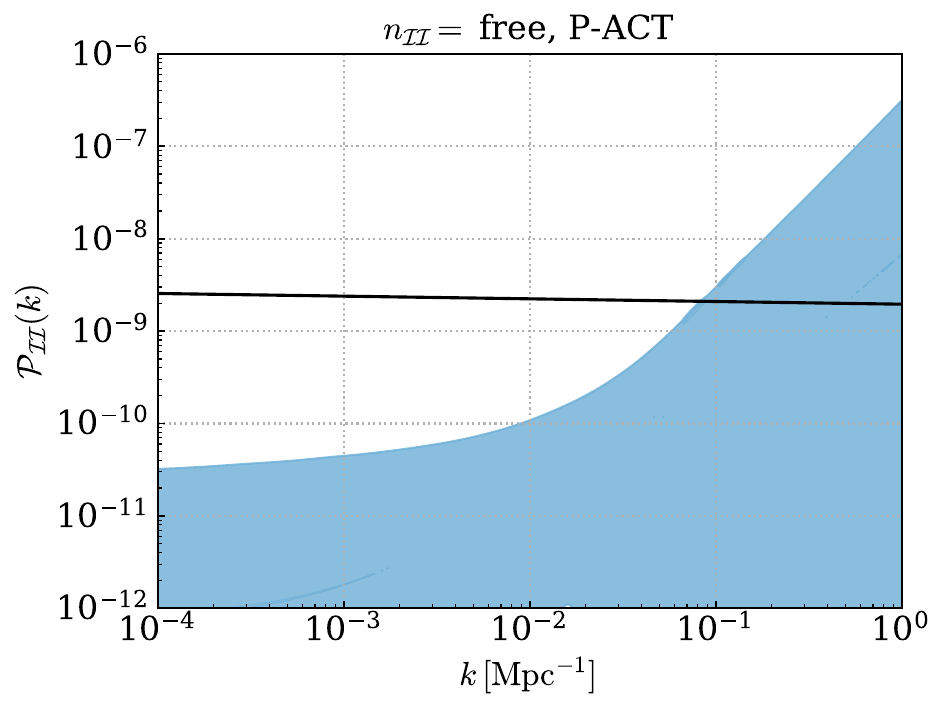}
    
    \includegraphics[width=0.49\columnwidth]{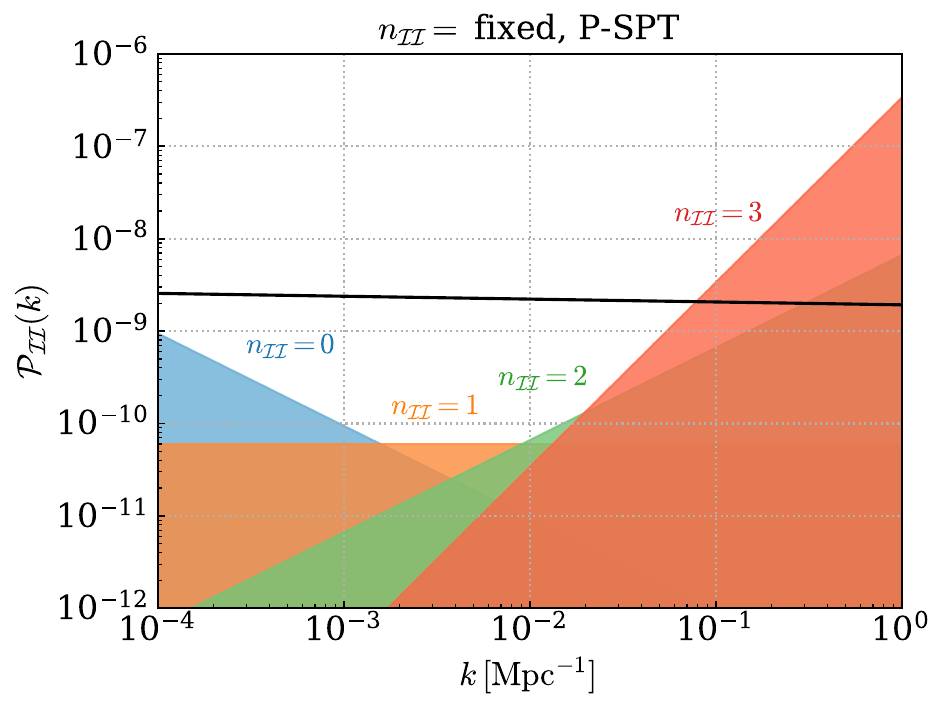}
    \hfill
    \includegraphics[width=0.49\columnwidth]{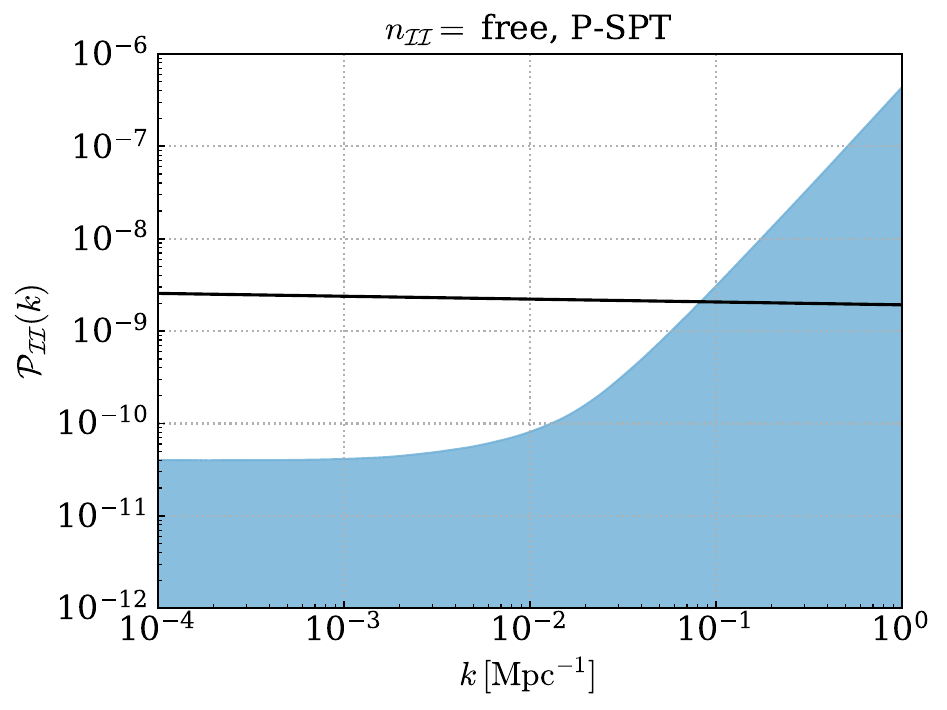}
    \caption{The 95\% C.L. constraints of the CDI power spectra for $\niso=0,1,2,3$ [left] and free $\niso$ [right]. The black lines are the best-fit curvature power spectra to the Planck,~P-ACT,~and P-SPT~datasets, respectively, assuming $\Lambda$CDM.}
    \label{fig:scale}
\end{figure}

\begin{figure}[ht]
    \centering
    \includegraphics[width=0.49\linewidth]{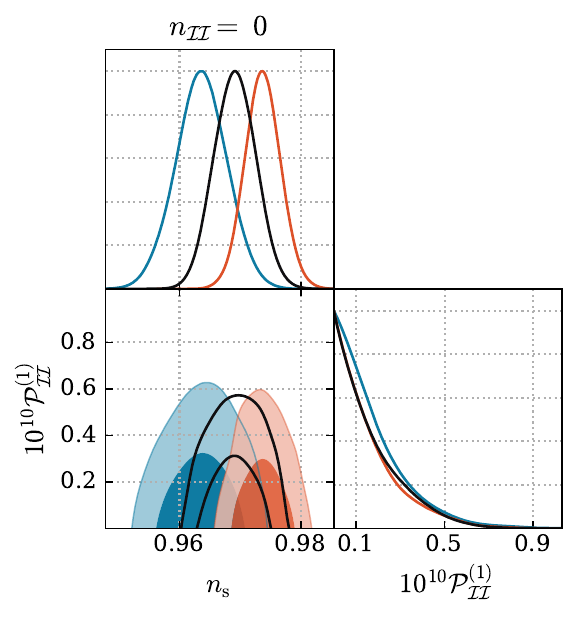}
    \hfill
    \includegraphics[width=0.49\linewidth]{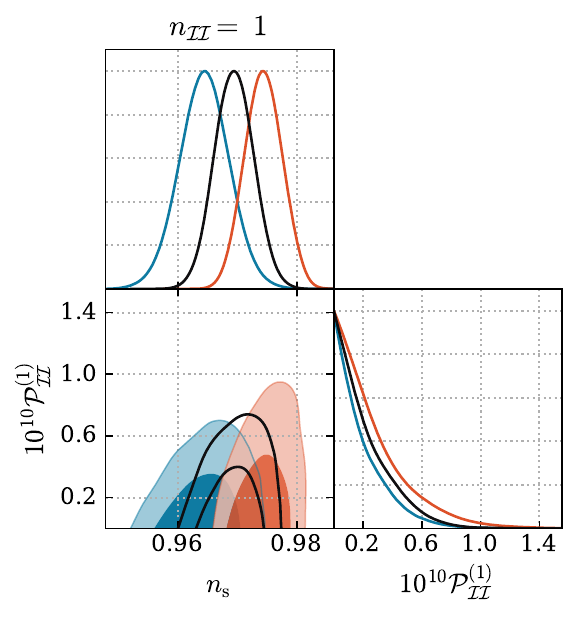}
    \includegraphics[width=0.49\linewidth]{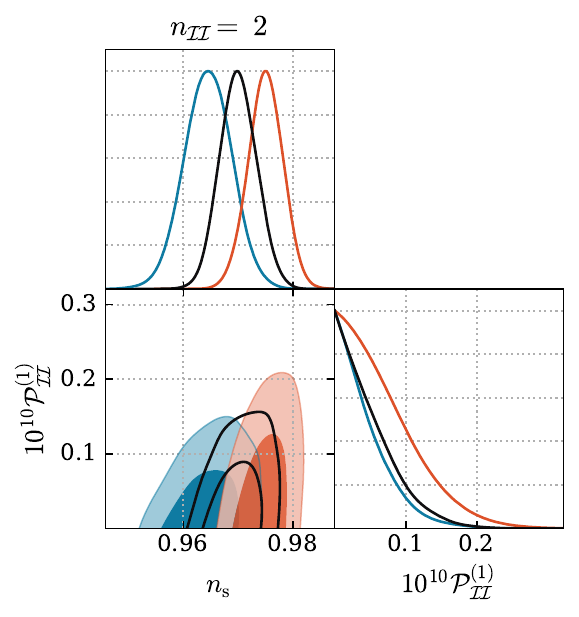}
    \hfill
    \includegraphics[width=0.49\linewidth]{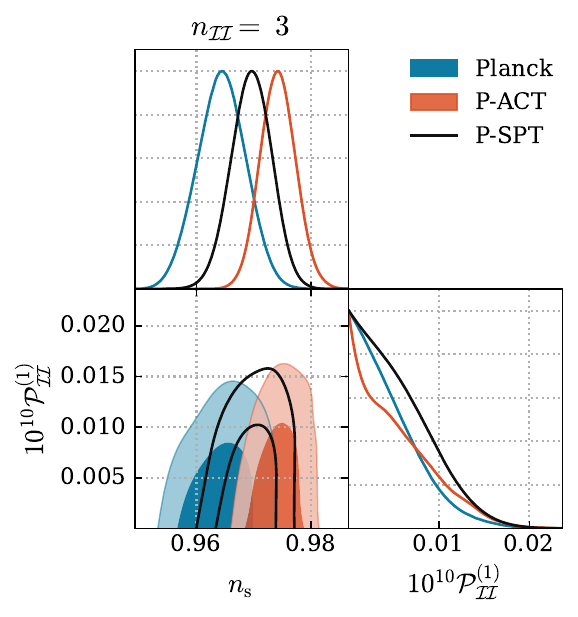}
    \caption{1D and 2D posterior distributions of $n_{\rm s}$ and $\PI$ for $\niso=0$ [top left], $\niso=1$ [top right], $\niso=2$ [bottom left], and $\niso=3$ [bottom right]. The 2D contours contain 68\% and 95\% of the probability.}
    \label{fig:ns_fixed}
\end{figure}

\begin{figure}[ht]
    \begin{center}
        \includegraphics[width=.54\columnwidth]{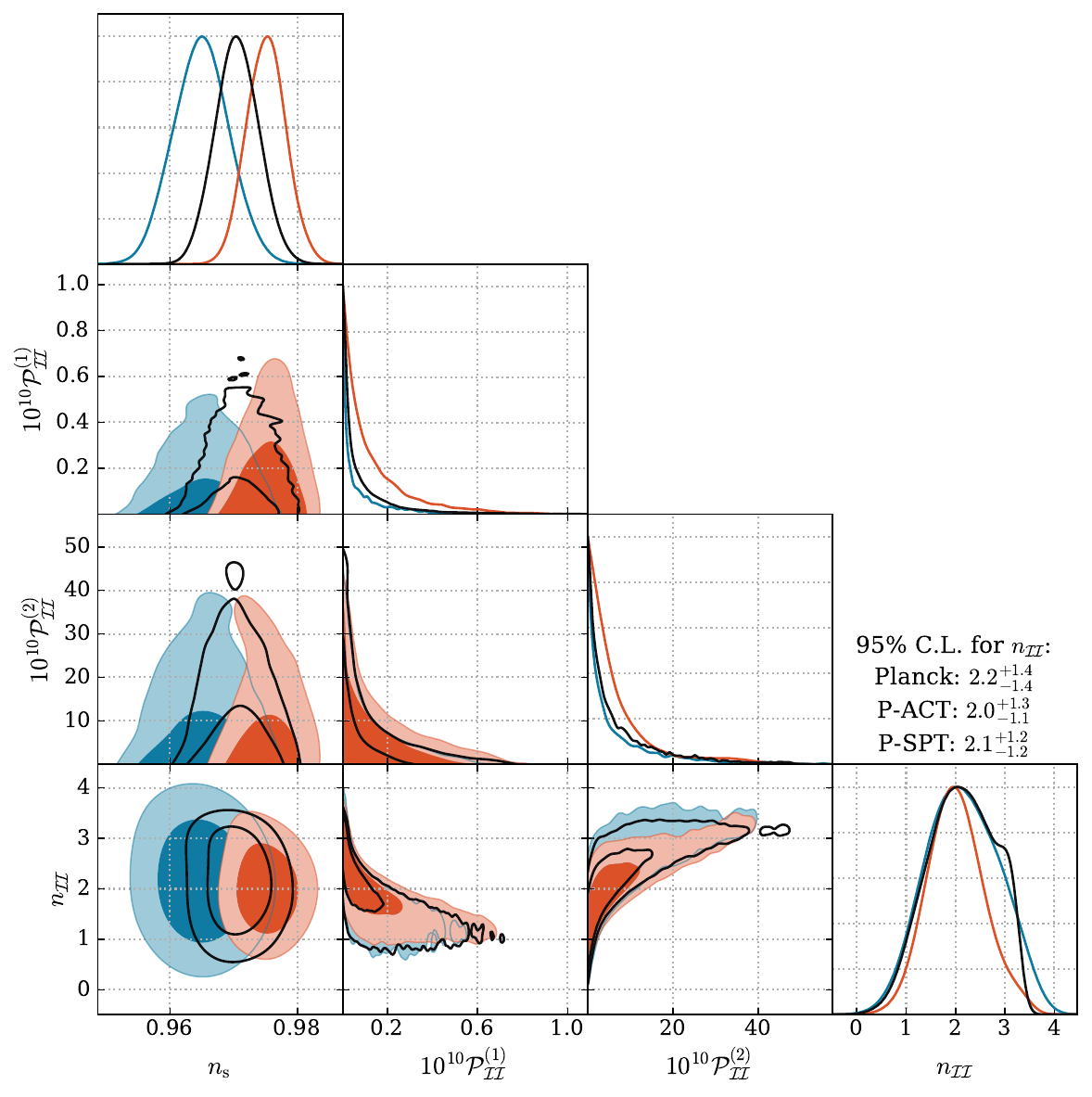}
        \hfill
        \includegraphics[width=.45\columnwidth]{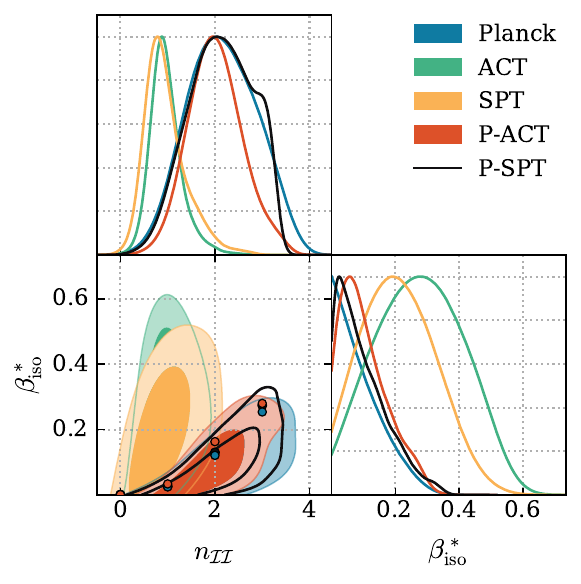}
    \end{center}
    \caption{\label{fig:constraints} 
    [Left] From our free $\niso$ analysis, 1D and 2D posterior distributions of the isocurvature parameters $\PI$, $\pii$, and the derived parameters $n_{\rm s}$ and $\niso$. The 2D contours contain 68\% and 95\% of the probability. [Right] 1D and 2D posterior distributions of the derived parameters $\biso^*$ and $\niso$. The points show the 95\% C.L. limits on the isocurvature fraction $\biso^*$ for given $\niso=0, 1, 2, 3$.}
\end{figure}

We express the isocurvature constraints for the fixed $\niso$ cases in several different ways. The left panels in Figure~\ref{fig:scale} show the 95\% C.L. upper bounds on the CDI power spectra as a function of $k$. Figure~\ref{fig:ns_fixed} shows the 1D and 2D posteriors for the curvature spectral index, $n_{\rm s}$, and $\PI$.  Lastly, the lower right panel of Figure~\ref{fig:constraints} shows the 95\% C.L. upper bounds on $\biso^*=\biso(k_*)$ as points plotted for each $\niso$.


From these constraints, we see that, in general, the P-ACT and P-SPT results do not offer much improvement compared to the Planck-only constraints. In fact, for flat- and blue-tilted spectra, the CDI amplitudes are slightly less constrained with P-ACT and P-SPT than Planck alone. In all cases, P-SPT offers slightly stronger constraints compared to P-ACT. The similarity between different constraints could be understood from Figure~\ref{fig:power_spectrum}. We can see that for \textit{TT} and \textit{EE}, the CDI power spectra decay much faster than the adiabatic spectra at small scales ($\ell \gtrsim 100$), which is solely due to the shape of the transfer functions. So, even though P-SPT and P-ACT extend to much smaller scales than Planck in temperature and polarization, on the scales that their constraining power is most relevant, the CDI transfer functions have already fallen off significantly. This explains why the addition of ACT or SPT does not offer much more constraining power for the CDI models, except perhaps for extremely blue-tilted cases of $\niso>3$.

For the $\niso=0$ case, P-ACT and P-SPT constraints are slightly better than the Planck constraints. While we keep $\niso=0$ as a benchmark, the extreme red-tilted spectrum is less motivated both theoretically and observationally.

However, for the flat and blue spectra, the constraints from P-ACT and P-SPT are mildly looser compared to Planck. This is mainly due to correlations between the curvature and isocurvature parameters. We find that the constraints on $\omega_b$, $\logA$, and $n_{\rm s}$ (the latter two of which are derived from $\pr$ and $\prr$) shift with each dataset, with the shifts being most noticeable in $n_{\rm s}$, as listed in App.~\ref{app:full_constraints}. For all isocurvature models, Planck gives the smallest values of these parameters while P-ACT gives the largest, and the P-SPT values lie between the other two datasets' values. This trend can be seen for all of the isocurvature models as well as for $\Lambda$CDM, see e.g.~\cite{ACT:2025fju}.\footnote{Note that in~\cite{SPT-3G:2025bzu}, their P-SPT combination uses all multipoles of the Planck data. Thus, the curvature parameters they report are shifted closer to the Planck values.} Furthermore, there is a slightly positive correlation between $n_{\rm s}$ and $\PI$, as seen in Figure~\ref{fig:ns_fixed} for fixed $\niso\geq 1$.  This correlation is due to the fact that for values of $n_{\rm s}$ closer to 1 (i.e. a flatter adiabatic spectrum), the curvature amplitude at large scales (such as $k_1=0.002~\Mpc^{-1}$, or $\ell\approx 28$) necessarily decreases compared to a redder spectrum. This allows for slightly larger isocurvature power at large scales. Planck gives the smallest $n_{\rm s}$ value and therefore the most stringent upper bounds on $\PI$ (and therefore on $\pii$ and the other derived isocurvature parameters). P-ACT gives the largest $n_{\rm s}$ value and therefore the loosest upper bounds on isocurvature parameters. P-SPT gives intermediate values between Planck and P-ACT. We also see that as $\niso$ increases from 1 to 3 and the isocurvature spectrum becomes more blue tilted, yielding less power at larger scales, the CDI constraints from P-ACT and P-SPT shift closer to the Planck values due to the datasets' increased constraining power on smaller scales.

Lastly, we comment that the particular choices of $\ell$ cuts we adopt to combine the datasets may also impact the constraints. For instance, in the P-ACT dataset, the Planck high-$\ell$ likelihoods are restricted to $\ell<1000$ for \textit{TT} since Planck and ACT have significant sky overlap in the discarded multipole range. However, the signal-to-noise in the ACT \textit{TT} data is larger than, or comparable to, the signal-to-noise in Planck until roughly $\ell \approx 1700$. Thus, this cut, while necessary to reduce correlations between Planck and ACT, could result in less precision on scales $\ell\approx 1000-1700$ in the P-ACT dataset which potentially worsen the constraints on the CDI spectra, especially for the $\niso=1$ and $\niso=2$ cases.

\subsection{Results for a Free Spectral Index}

Now, let us discuss the results for free $\niso$. The left panel of Figure~\ref{fig:constraints} shows a triangle plot for $n_{\rm s}$, $\PI$, $\pii$, and $\niso$, and the right panel shows a triangle plot for the derived parameters $\niso$ and $\biso^*=\biso(k_*)$ for all of the datasets, including ACT- and SPT-only results. The interpretation of the results for the free $\niso$ case is slightly different than that for the fixed $\niso$ cases because the constraints on small and large scales depend on the constraints on the spectral tilt. We see that the ACT and SPT-only results favor flat and red spectra (smaller $\niso$). This is because both ACT and SPT have extra constraining power on very small scales ($\ell>2000$) and less constraining power on large scales compared to Planck, allowing for a redder spectrum. Meanwhile, Planck has the most constraining power on large scales and thus prefers a bluer spectrum. We see that adding Planck to ACT or SPT data has the effect of pulling the spectrum more blue and the constraint much closer to the Planck-only one. This is sensible since, as we discussed for the fixed $\niso$ cases, the CDI transfer functions have mostly decayed by the scales relevant for ACT and SPT to offer the most constraining power.

However, there is still a preference for a slightly redder spectrum for P-ACT and P-SPT compared to Planck, which is consistent with the result reported in~\cite{ACT:2025tim}. As a result, the constraint on $\niso$ improves mildly with P-ACT and P-SPT. We find the constraint on the spectral index at 95\% C.L. from Planck is $\niso=2.2^{+1.4}_{-1.4}$, from P-ACT is $\niso=2.0^{+1.3}_{-1.1}$, and from P-SPT is $\niso=2.1^{+1.2}_{-1.2}$. The constraint on $\niso$ from P-SPT lies between the Planck and P-ACT constraints. An extremely blue-tilted spectrum with $\niso\gtrsim 4$ is ruled out at 95\% C.L.

In Figure~\ref{fig:constraints}, we also see the correlation between a redder spectrum with a larger constraint on $\biso^*$. This is due to the fact that the constraints at large scales are directly affected by those at small scales via the spectral tilt. For example, if the power at the smaller length scale $\pii$ is kept fixed and $\niso$ decreases, then the power at the larger scale $\PI$ will necessarily increase. The constraints on $\pii$ from Planck, P-ACT, and P-SPT are roughly similar, which could be seen from Table~\ref{tab:constraints}.
Therefore, because P-ACT and P-SPT favor a more red-tilted spectrum, their constraints on $\PI$ are slightly worse than the constraint from Planck. In contrast, Planck favors a more blue-tilted spectrum and thus the constraint on $\PI$ is tighter. Notice also that P-ACT prefers the reddest spectrum and the largest value of $\PI$, but also the largest value of $n_{\rm s}$, while the opposite is true for Planck. The P-SPT values are between those from the other two datasets, which further demonstrates the correlation between $n_{\rm s}$ and $\PI$.

\subsection{Estimates of Potential Future Experiments}

\begin{figure}[ht]
    \centering
    \includegraphics[width=\linewidth]{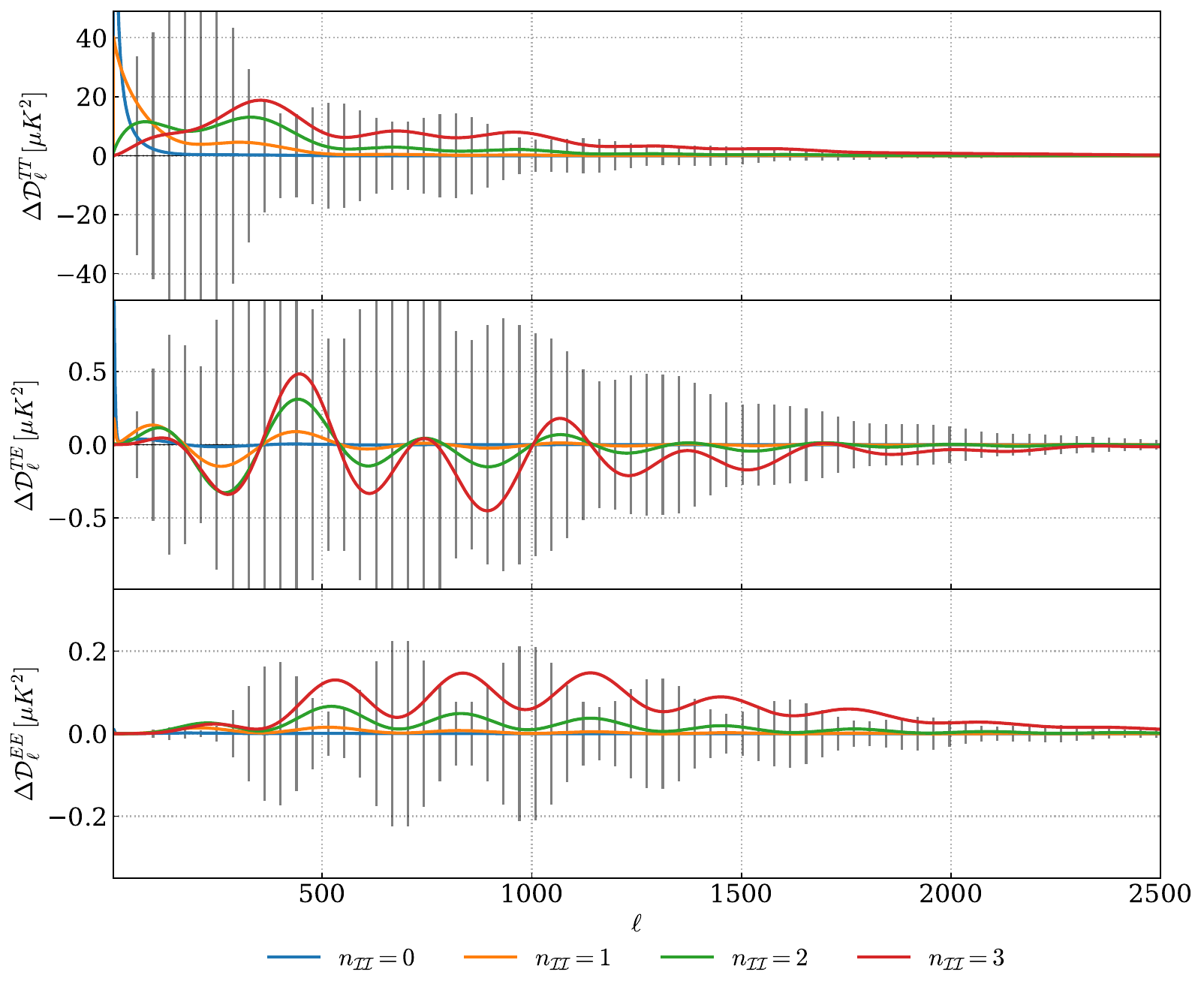}
    \caption{CMB power spectrum residuals of CDI models with
    respect to $\Lambda$CDM. The CDI models have cosmological parameters set to the same values as the $\Lambda$CDM baseline and $\PI$ set to the 95\% C.L. upper bound from Planck for each respective $\niso$. The gray error bars show the cosmic variance limit to represent the most optimal experimental precision. For a better visualization of how well the acoustic oscillations can be resolved, we bin the error bars using a bin width of $\Delta \ell=38$.}
    \label{fig:forecasts}
\end{figure}

In this subsection, we qualitatively assess the prospects for improving isocurvature constraints using future experimental datasets. Specifically, we estimate roughly whether the reduced uncertainties achievable by future CMB experiments could lead to tighter constraints on isocurvature perturbations.

We assume the best-fit $\Lambda$CDM model from our Planck dataset to be the ``true" model of our universe and consider the multipole range $\ell=2-2500$ probed by Planck. We then examine the residuals of the CMB power spectra for models with isocurvature perturbations with respect to the $\Lambda$CDM model. As a best-case scenario, we consider error bars arising solely from cosmic variance. This effectively corresponds to a perfect, noiseless experiment with full-sky coverage and no systematic uncertainties. Therefore, it represents the most optimistic scenario achievable by any future CMB experiment.

For better visualization, we bin the cosmic variance error bars. Since CDI perturbations are out of phase with adiabatic fluctuations, the residuals exhibit oscillatory behavior. In order to adequately resolve these oscillations, we choose a bin size of $\Delta \ell=38$. This corresponds to approximately $8-10$ data points per oscillation period of the CDI models we work with.

Figure~\ref{fig:forecasts} shows the CMB power spectrum residuals of CDI models with fixed spectral index values, $\niso=\{0,1,2,3\}$. For these CDI models, cosmological parameters are set to the same values as the $\Lambda$CDM baseline, with $\PI$ set to the $95\%$ C.L. upper bound obtained from our Planck constraints. For blue-tilted CDI models i.e. $\niso=\{2,3\}$, we see scale-independent, positive excess components in the residuals of both \textit{TT} and \textit{EE} CMB spectra. Since these excesses are discernible from the $\Lambda$CDM baseline across a wide range of scales, particularly in \textit{EE}, future experiments with reduced uncertainties are expected to place tighter constraints on these models. In particular, while $\niso = 2$ represents a borderline case, the $\niso = 3$ model exhibits a more pronounced excess, and is therefore expected to be more strongly constrained by future experiments. 

\section{Implications for QCD Axion Dark Matter}
\label{sec:axionconstraints}
In this section, we will discuss the theoretical implications of the CMB isocurvature constraints for pre-inflationary QCD axion DM. We start with the flat spectrum case, whose implications are already well known. We then discuss the less-explored scenario with a blue-tilted spectrum, and present its feasible parameter space.

\subsection{Flat Spectrum}

\begin{figure}[ht]
    \begin{center}\includegraphics[width=0.6
    \textwidth]{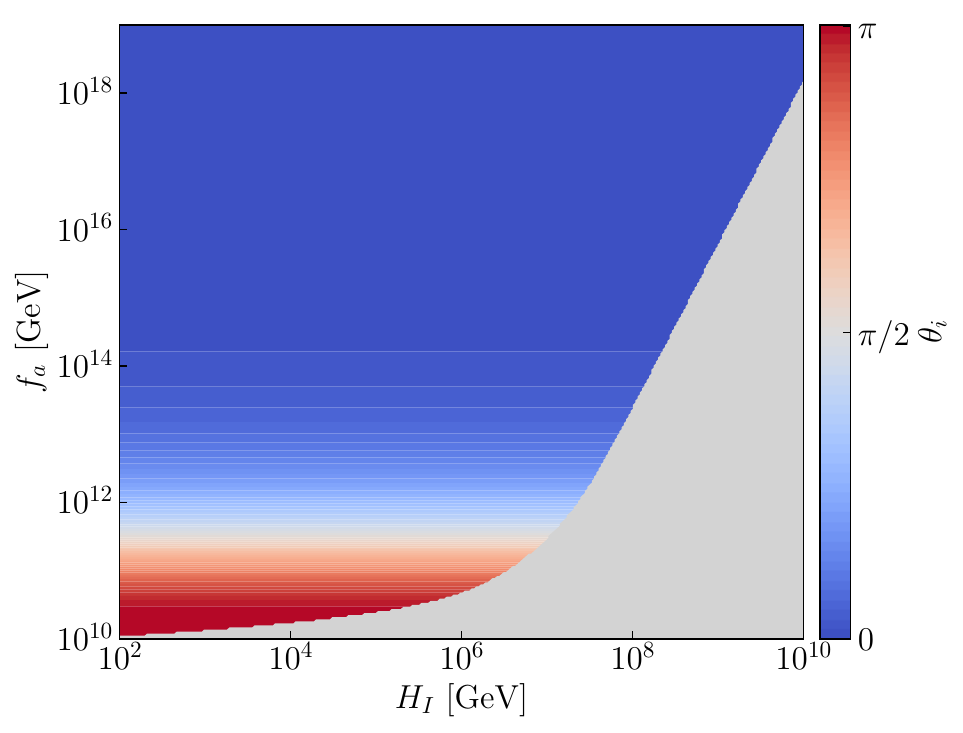}
    \end{center}
    \caption{Constraints on pre-inflationary QCD axion DM in the $(H_I, f_a)$ plane. The gray shaded region is excluded by the non-observation of CDI by P-ACT. Here, the QCD axion is the entire DM with $\Omega_a = \Omega_d$. We also show the corresponding $\theta_i$ for each given $f_a$ in the color bar.
  \label{fig:axion1} }
		\end{figure}

If $\niso=1$, we could apply Eqs.~\eqref{eq:Ai} and \eqref{eq:abundance} to find out the allowed parameter space in the $(H_I, f_a)$ plane, assuming $f_I=f_a$. The results are presented in Figure~\ref{fig:axion1}, with the P-ACT bound. We solve $\theta_i$ for a given $f_a$ using Eq.~\eqref{eq:abundance} and requiring $\gamma=1$ (DM is entirely the QCD axion). For smaller $f_a$'s, $\theta_i$ increases until it approaches its maximum value $\pi$ at $f_a\sim10^{10}~\GeV$. From the plot, one could see that when QCD axion DM is the dominant component of DM and $\theta_i$ is natural ($\theta_i \sim 1$), the inflationary Hubble scale has to be small, $H_i \lesssim 10^7~\GeV$. Even if we allow a unnaturally tiny $\theta_i$ so that the axion decay constant gets close to the Planck scale $10^{18}~\GeV$, $H_I$ still has to be below $10^{10}~\GeV$ to satisfy the CDI constraint. This is the (in)famous QCD axion DM isocurvature problem: DM being mainly the QCD axion is incompatible with high-scale inflation models. From the future observational perspective, this implies that an observation of  primordial $B$-modes by next-generation CMB experiments--probing a tensor-to-scalar ratio of order $10^{-3}$ and implying $H_I \gtrsim 10^{13}~\GeV$---would exclude the QCD axion as the dominant component of DM in the minimal scenario. 

In Figure~\ref{fig:axion1}, the upper bound on $H_I$ decreases roughly linearly when $f_a$ decreases, when $\theta_i \lesssim 1$. But when $\theta_i$ is close to $\pi$ in the large misalignment limit, or equivalently $f_a$ gets closer to $10^{10}~\GeV$, the upper bound on $H_I$ gets significantly stronger, as first pointed out in Ref.~\cite{Kobayashi:2013nva}. More specifically, for $f_a = 2.5 \times 10^{10}~\GeV$, $H_I \lesssim 10^5~\GeV$; and when $f_a = 1.3 \times 10^{10}~\GeV$, $H_I \lesssim 10^3~\GeV$. This could be understood from the $\partial \ln \Omega_a/\partial \theta_i$ factor in Eq.~\eqref{eq:Ai}, which shows that isocurvature could be significantly enhanced since $\Omega_a$ is sensitive to the initial misalignment angle near the hilltop region of $\theta_i\sim \pi$.

\subsection{Blue-tilted Spectrum}  
\label{subsec:bluespec}

\begin{figure}
    \centering
    \includegraphics[width=0.6\textwidth]{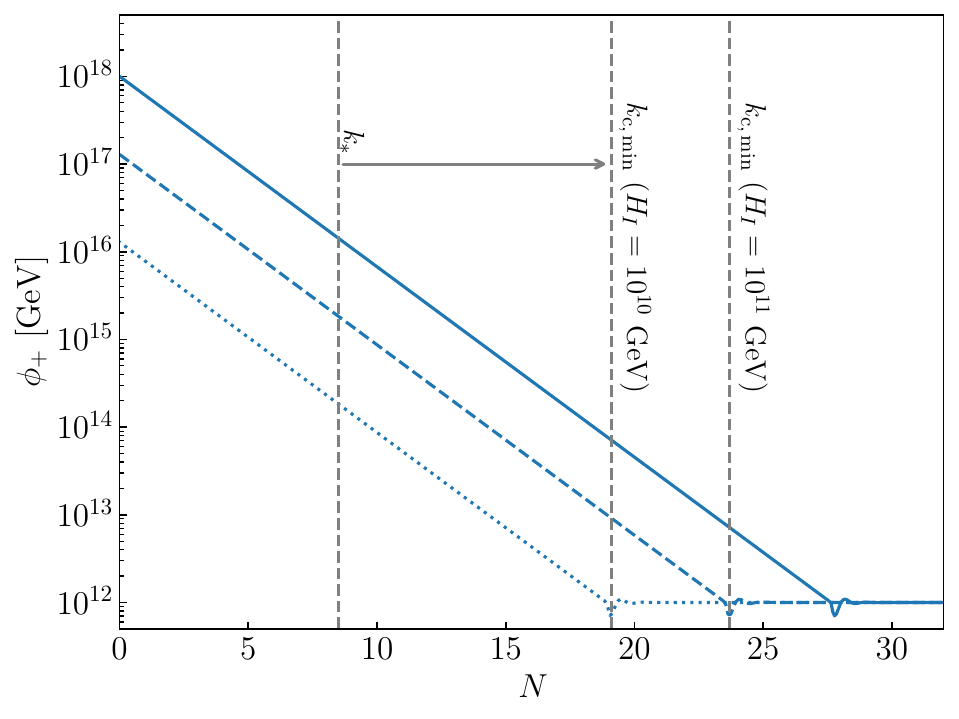}
    \caption{The schematic plot of the field $\phi_+$'s evolution during inflation as a function of the number of e-folds $N$. The shown benchmarks correspond to the case of a QCD axion with $\theta_i\sim 1$. The benchmarks correspond to $\phi_+ \simeq f_a \simeq 10^{12}~\GeV$ at late times of inflation when $\phi_+$ stops evolving. In all cases, $\phi_+$ starts with a large initial value and scales as $e^{-N(\niso-1)/2}$ until its value is set at $f_a =10^{12}~\GeV$ at wavenumber $k_{c}$. Near the horizon crossing of $k_c$, $\phi_+$ may undergo non-trivial dynamic relaxation processes and introduce characteristic features. The $N=0$ ``initial condition" refers to a benchmark scale of $\ki=10^{-5}~\Mpc^{-1}$, and the leftmost vertical line represents the scale $k_*=0.05~\Mpc^{-1}$. The other two vertical lines on the right correspond to the minimum transition scale $k_{c, \rm{min}}$ for given Hubble scales to satisfy the isocurvature bounds. }
    \label{fig:bluetiltcartoon}
\end{figure}

One representative model leading to a blue-tilted primordial isocurvature spectrum depends on non-trivial dynamics of the supersymmetric PQ scalar fields during inflation~\cite{Kasuya:2009up}. Due to supergravity effects, the PQ fields acquire large Hubble-induced mass terms during inflation, which drive their non-trivial evolution. In the most relevant direction in the field space, the field, denoted as $\phi_+$, could start with a large value which could be as high as the Planck scale, associated with a small isocurvature power. It will then decay exponentially for a short period during inflation, and finally settle down into the minimum of the potential with its value being the axion decay constant today $f_a$, which gives rise to a larger isocurvature power. As a result, the primordial axion isocurvature spectrum is blue tilted at large length scales, and scale invariant at smaller length scales. The isocurvature spectral index is a function of $\mathcal{O}(1)$ parameters in the model. 

A schematic plot of $\phi_+$ evolution is shown in Figure~\ref{fig:bluetiltcartoon} for the $\niso=2$ case. In the figure, the wavenumber of the mode that starts to exit the horizon at each e-fold $N$ during inflation increases as $N$ increases. At a characteristic cutoff scale, $\kc^{-1}$, $\phi_+$ stops evolving and settles down to $f_a$. The cutoff scale is where the isocurvature spectrum transitions from a blue-tilted one to a flat one. Following discussions in~\cite{Kasuya:2009up}, when the field $\phi_+$ becomes dynamic during inflation, it behaves as a scale-dependent $f_I$ in Eq.~\eqref{eq:Ai} and scales as $e^{-N(\niso-1)/2}$. Equivalently, tracking backward in time from the moment of transition, we have
\begin{equation}
    \phi_+(k) \simeq f_a \bigg(
    \frac{\kc}{k}\bigg)^{\frac{\niso-1}{2}} ,~k<\kc~,
    \label{eq:kc1}
\end{equation}
where $\phi_+(k)$ stands for the $\phi_+$ field value when the comoving mode of $k$ exits the horizon.
Note that in this scenario, we assume that the evolution after inflation to be the standard misalignment process of QCD axion DM. Consequently, the relation between $\Omega_a$ and $f_a$ in Eq.~\eqref{eq:abundance} remains unchanged. For instance, assuming $\gamma=1$ and $\theta_i=1$, $f_a \approx 10^{12}~\GeV$. 

\begin{figure}[ht]
    \centering
    \includegraphics[width=0.495\linewidth]{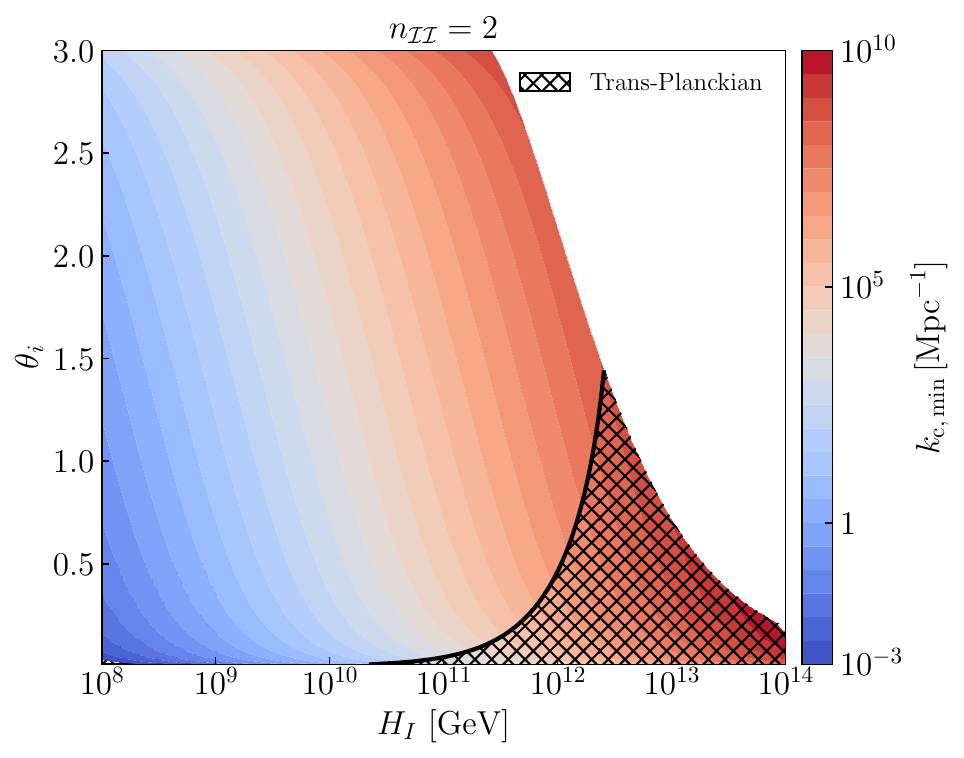}
    \includegraphics[width=0.495\linewidth]{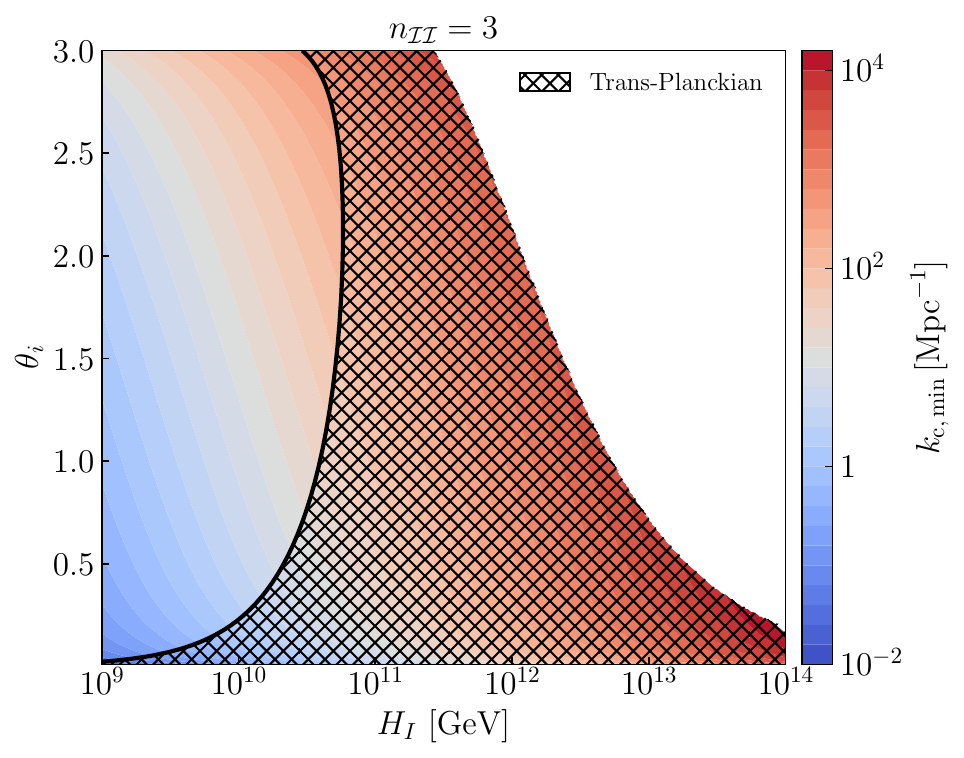}
    \caption{The smallest possible scale for spectrum transition, $\kcm$, in the blue-tilted model. [Left] $\niso=2$. [Right] $\niso=3$. Here, we assume the post-inflationary dynamics are the same as the standard case in Sec.~\ref{sec:axion}. Therefore, $f_a$ is uniquely determined by $\theta_i$ according to Eq.~\eqref{eq:abundance}. At a given $f_a$ and $\theta_i$, if the transition appears at even larger length scales (or equivalently, smaller $\kcm$) than indicated in the plot, it will violate the P-ACT isocurvature bounds. The upper right blank region is associated with $H_I>2\pi f_a$ and the pre-inflationary axion scenario no longer applies. The meshed region stands for the case where the initial value, $\phi_+(\ki)$, becomes trans-Planckian. We choose $\ki$ to be $10^{-5}~\Mpc^{-1}$. For $\kcm$ $\lesssim 0.1~\Mpc^{-1}$, the cutoff feature affects the CMB directly, and the exact numbers become highly model-dependent. }
    \label{fig:bluetilted}
\end{figure}

We could use the constraint in Table~\ref{tab:constraints} to estimate the upper bound on the ratio $H_I/\phi_+({k_\ast})$ at a reference scale, e.g. $k_\ast=0.05~\Mpc^{-1}$. Since $\phi_+ \gg f_a$ at early stages of inflation, the isocurvature constraint on $H_I/\phi_+(k_\ast)$ instead of $H_I/f_a$ greatly relaxes the constraint on $H_I$. Assuming $\gamma = 1$ and $\theta_i$ is away from the hilltop region so that $\partial \ln \Omega_a/\partial \theta_i\approx 2/\theta_i$, and using Eq.~\eqref{eq:kc1} and the P-ACT results (the weakest constraints for a given $\niso \geq 1$) in Table~\ref{tab:constraints}, we find that at $k_\ast=0.05~\Mpc^{-1}$:
\begin{equation}
    \frac{H_I}{\phi_+(k_\ast)\theta_i} \simeq \frac{H_I}{f_a\theta_i}\bigg(\frac{k_\ast}{\kc}\bigg)^{\frac{1}{2}} <  6.4\times 10^{-5}~,~\niso=2~ , 
    \label{eq:kcn2}
\end{equation}
and
\begin{equation}
    \frac{H_I}{\phi_+(k_\ast)\theta_i} \simeq \frac{H_I}{f_a \theta_i}\bigg(\frac{k_\ast}{\kc}\bigg) < 9.0\times 10^{-5}~,~\niso=3~. 
    \label{eq:kcn3}
\end{equation}
From the equations above, we see that for a given $H_I$, there exists a lowest
$\kcm$, which saturates the P-ACT bound. This is illustrated by the two rightmost vertical dashed lines in Figure~\ref{fig:bluetiltcartoon}, corresponding to $\kcm$'s for $H_I=10^{10}~\GeV$ and $H_I = 10^{11}~\GeV$, respectively.\footnote{When $\theta_i$ is close to $\pi$, the anharmonic correction of Eq.~\eqref{eq:abundance} is taken into account.} The length scale $\kcm^{-1}$ corresponds to the largest possible length scale where the transition between a blue-tilted and a flat spectrum occurs.\footnote{Depending on dynamical details, the transition at the cutoff scale could be approximately smooth or oscillatory, introducing non-trivial phenomena such as primordial features in the isocurvature spectrum~\cite{Chung:2021lfg} or a featured gravitational wave signal~\cite{Garcia:2025wmu}, that may be tested with future gravitational wave interferometers~\cite{LISACosmologyWorkingGroup:2024hsc,LISACosmologyWorkingGroup:2025vdz}.} The $\kcm$ value as a function of $H_I$ and $\theta_i$ is shown in Figure~\ref{fig:bluetilted} for both $\niso=2$ (left panel) and $\niso=3$ (right panel). In the blank area in the upper right region of each panel, $H_I> 2\pi f_a$, and the PQ symmetry is restored during later times of inflation. Thus, there is no axion present, and the isocurvature constraint no longer applies. For both $\niso=2$ and 3, as $H_I$ increases, $\kcm$ is pushed to higher values to satisfy the CMB constraint. Equivalently, the transition from the blue-tilted to the flat spectrum has to happen at smaller length scales when $H_I$ increases. Conversely, for sufficiently low $H_I$, the corresponding $\kcm$ will be lower than $\mathcal{O}(0.1)~\Mpc^{-1}$ and the corresponding cutoff features affect the CMB observables directly. The constraints on $\mathcal{P}_\mathcal{II}^{(1)}$ obtained in Sec.~\ref {sec:results} no longer apply exactly. Therefore, the corresponding $\kcm$ values in the lower-left corners of Figure~\ref{fig:bluetilted} become model-dependent and are only shown as reference values. 

Note that $k_*=0.05~\Mpc^{-1}$ is only an intermediate reference scale widely used in CMB analysis. 
In the blue-tilted model with the $H_I$ constraint relaxed, $\phi_+$ starts evolving from an initial value at a scale $\ki \ll k_\ast$, $\phi_+(\ki)$. We take the benchmark value of $\ki=10^{-5}~\Mpc^{-1}$, which conservatively corresponds to the largest length scale that CMB experiments could probe. We could rewrite the evolution of $\phi_+$ from $\ki$ as $ \phi_+(k)\simeq \phi_+(\ki) (\ki/k)^{(\niso-1)/2}$ when $\ki < k < k_c$. The initial condition has to satisfy
\begin{eqnarray}
    \phi_+(\ki)= \phi_+(k_\ast)\bigg(\frac{\ki}{k_\ast}\bigg)^{-\frac{\niso-1}{2}} \gg \phi_+(k_\ast)\gg H_I~,
\end{eqnarray}
so that the axion is present during inflation. 
The lower limit of $\phi_+(k_\ast)$ is given by the P-ACT constraint in Eq.~\eqref{eq:kcn2} and~\eqref{eq:kcn3}, for a given $H_I$ and $\theta_i$ and assuming that axion is the only component of DM ($\gamma=1$). Such a large initial value of $\phi_+$ may become trans-Planckian for high $H_I$. In Figure~\ref{fig:bluetilted}, we show the region where $\phi_+(\ki)$ becomes trans-Planckian with the P-ACT constraint as meshed regions. Those regions are theoretically problematic with trans-Planckian field excursion. Even taking into account of this theoretical constraint, we could see from Figure~\ref{fig:bluetilted} that the maximum $H_I$ values allowed in the blue-tilted cases, $3\times 10^{12}(6\times 10^{10})~\GeV$ for $\niso=2(3)$, are still significantly higher than the $10^{7}~\GeV$ shown in Figure~\ref{fig:axion1} for the $\niso=1$ case, assuming $\gamma=1$ and $\theta_i = 1$.\footnote{Bounds on $H_I$ could be further relaxed when considering more complicated dynamics of $\phi_+$, which is beyond the scope of this work.}

\section{Conclusions}
\label{sec:con} 

Pre-inflationary QCD axion dark matter could lead to observable isocurvature and serves as a highly motivated target for CMB measurements. In this article, we present the most updated constraints on axion CDI, with various combinations of the latest CMB data. We implement nested sampling analyses over a wide range of fixed CDI spectrum indices, ranging from red tilted to blue tilted, as well as an analysis with a free CDI index. 

For a given primordial axion isocurvature index, it turns out that Planck data with the full likelihood (instead of the {\tt lite} version) still gives the strongest constraint on the isocurvature amplitude. The combined datasets, P-ACT and P-SPT, give comparable bounds without further improvements. This is mainly due to the fact that CDI transfer functions decay faster than the curvature ones and the $\ell$ range well covered by Planck remains as the most sensitive probe to axion isocurvature. In addition, mild shifts in the preferred values of curvature parameters, i.e. $n_{\rm s}$, among different CMB datasets also result in slightly looser constraints on the CDI amplitudes. Yet the combined datasets, P-ACT and P-SPT, still lead to moderate improvements on constraining the primordial CDI spectral index, $\niso$, when it is a free parameter, driven by the precise large $\ell$ measurements of ACT and SPT. For both P-ACT and P-SPT, the central values of the $\niso$ posteriors shift closer to $\sim 2$ from the Planck value of 2.2, with slightly narrower error bars. We also work out the theoretical implications for QCD axion DM scenarios. In particular, we work out the feasible parameter space for scenarios with primordial blue-tilted axion isocurvature spectra, combining both the CMB CDI constraint and theoretical limit on the trans-Planckian field range.

Future CMB measurements will improve polarization measurements and push their uncertainties closer to the cosmic variance limits. As an illustration, we compare currently allowed isocurvature residuals with cosmic variance limits, and find that indeed those future CMB measurements, in particular, the polarization modes, could be more sensitive to small deviations generated by CDI. In addition, large-scale structure measurements probe density perturbations toward the small-scale end of the range accessible to the CMB, and often extend to much smaller scales. It would be also interesting to study the constraints coming from large-scale structure, in particular for very blue-tilted spectra. 

An additional test of CDI could be to search for oscillatory features in the primordial spectrum, such as the models proposed in~\cite{Chen:2023txq}. However, without a detection of the CDI amplitude, constraints on the feature parameters (the oscillation amplitude, frequency, and phase) would be less informative. However, if future high-precision CMB experiments do make a detection of CDI, then primordial feature models may prove to be an interesting avenue for further investigation.

\section*{Acknowledgements}
We thank Lennart Balkenhol, Nicola Barbieri, Jens Chluba for very useful correspondence on SPT, ACT, and CMB spectral distortions, respectively. JF and PS are supported by the NASA grant 80NSSC22K081 and the DOE grant DE-SC-0010010. MB has received funding from the European Union’s Horizon 2020 research and innovation program under the Marie Skłodowska-Curie grant agreement No 101205460. This work was conducted using computational resources and services at the Center for Computation and Visualization (CCV), Brown University and the Faculty of Arts and Sciences Research Computing (FASRC), Harvard University.

\appendix

\section{Relations between Two-Scale and Power-Law Parameterizations}
\label{app:notations}

The mapping between the single power-law parametrization in Eq.~\eqref{eq:nidefinition} and Eq.~\eqref{eq:beta_iso} and the two-scale parametrization in Eq.~\eqref{eq:two-scale} is given by 
\begin{align}
    n_{\rm s}=&1+\frac{\ln(\pr)-\ln(\prr)}{\ln(k_1)-\ln(k_2)}\, ,\notag\\
    A_{\rm s}=&\exp\left[\frac{\ln(k_1)\,\ln(\prr) - \ln(k_2)\,\ln(\pr)+\ln k_* \,\left(\ln(\pr)-\ln(\prr)\right)}{\ln(k_1)-\ln(k_2)}\right]\, ,\notag\\
    \niso=&1+\frac{\ln(\PI)-\ln(\pii)}{\ln(k_1)-\ln(k_2)}\, ,\notag\\
    \fiso^2=&A_{\rm s}^{-1}\exp\left[\frac{\ln(k_1)\,\ln(\pii) - \ln(k_2)\,\ln(\PI)+\ln k_* \,\left(\ln(\PI)-\ln(\pii)\right)}{\ln(k_1)-\ln(k_2)}\right] \, .\label{eq:map_between_parameterizations}
\end{align}
Using these equations, one can adopt the two-scale approach and derive constraints on the theory-oriented power-law parameterization in post-processing. 

\section{Comparison of CDI Constraints with Different Planck Likelihoods}
\label{app:likelihoods}
In this appendix, we comment on the differences in likelihoods between our results and those used by the ACT collaboration in their CDI analysis~\cite{ACT:2025fju,ACT:2025tim}. As mentioned in Section~\ref{subsec:data}, the P-ACT dataset that is used by the ACT collaboration uses the Planck {\tt plik\_lite} likelihood, rather than the {\tt CamSpec} likelihood that we use in our analysis. {\tt plik\_lite} is a CMB-only likelihood and includes marginalization over foregrounds and nuisance parameters~\cite{Planck:2019nip}. {\tt plik\_lite} is useful for fast computations within the $\Lambda$CDM framework, but it is best practice to avoid this likelihood when analyzing extended cosmological models and instead use a full high-$\ell$ likelihood, such as {\tt CamSpec}, that requires fitting for nuisance parameters. The ACT collaboration uses {\tt plik\_lite} in their analyses rather than a full Planck high-$\ell$ likelihood because of poorer agreement between the ACT and {\tt NPIPE} spectra~\cite{ACT:2025fju}. In our analysis, we use the full PR4 version of the {\tt CamSpec} likelihood, but keep the same cuts in the Planck likelihood at $\ell\in[30,1000]$ for \textit{TT} and $\ell\in[30,600]$ in \textit{TE}/\textit{EE} as the ACT collaboration had done. This choice was made to achieve better constraining power on extended cosmological models with a full high-$\ell$ Planck likelihood while still minimizing the discrepancy between the ACT and PR4 spectra.

\begin{table}[ht]
    \centering
    \begin{tabular}{|l|l|l|l|l|l|}
        \hline
        Model and Likelihood & $\PI$ & $\pii$ & $100~\biso^{(1)}$ & $100~\biso^{(2)}$ & $\niso$ \\
        \hline \multicolumn{6}{|l|}{ $\niso=1$} \\
        \hline 
        
        {\tt plik\_lite} & $<9.34\times 10^{-11}$ & $<9.34\times 10^{-11}$ & $<3.83$ & $<4.34$ & $\hdots$ \\
        {\tt CamSpec} & $<5.30\times 10^{-11}$ &  $<5.30\times 10^{-11}$ & $<2.21$ & $<2.51$ & $\hdots$ \\
        ACT collaboration & $<9\times 10^{-11}$ & $<9\times 10^{-11}$ & $<3.7$ & $<4.2$ & $\hdots$ \\
        \hline \multicolumn{6}{|l|}{ $\niso$ free} \\
        \hline 
        
       {\tt plik\_lite} & $<4.27\times 10^{-11}$ & $<6.53\times 10^{-9}$ & $<1.78$ & $42_{-40}^{+36}$ & $2.4_{-1.1}^{+1.2}$ \\
        {\tt CamSpec} & $<3.63\times 10^{-11}$ &  $<2.84\times 10^{-9}$ & $<1.54$ & $<58.2$ & $2.2_{-1.4}^{+1.4}$ \\
        ACT collaboration & $<5\times 10^{-11}$ & $<5.9\times 10^{-9}$ & $<2.0$ & $41_{-41}^{+34}$ & $2.4_{-1.2}^{+1.1}$ \\
        \hline
    \end{tabular}
    \caption{Constraints on $\niso=1$ and free $\niso$ CDI models using the Planck dataset. The results use either the {\tt plik\_lite} or the {\tt CamSpec} likelihood for high-$\ell$ Planck data. Results from the ACT collaboration~\cite{ACT:2025fju} for the Planck dataset are also shown for comparison. The results are reported at 95\% C.L.}
    \label{tab:pliklite}
\end{table}

As a result of choosing a different Planck likelihood for high-$\ell$ data, we find that our constraints on the CDI spectra are tighter than those reported by the ACT collaboration~\cite{ACT:2025tim} for $\niso=1$ and free $\niso$. The impact on the constraints due to this choice of likelihood is most notable for our Planck-only results but is not as dramatic for the P-ACT dataset (see Table~\ref{tab:constraints}). As a sanity check, we also performed the same analysis as~\cite{ACT:2025tim} for the Planck dataset with the {\tt plik\_lite} likelihood and obtained very similar constraints on the CDI spectra for $\niso=1$ and free $\niso$ as the collaboration. We list these results in Table~\ref{tab:pliklite}.

\section{Constraints from CMB Spectral Distortions}
\label{app:SDs}

In this appendix, we discuss constraints on CDI perturbations from CMB spectral distortions (SD), which are produced by the Silk-damping of photons. CMB spectral distortions are tightly constrained by FIRAS, which provides the following 95\% C.L. constraints on $\mu$- and $y$-type distortions:
\begin{align}
    \lvert\mu\rvert<&~ 9.5\times10^{-5}~,\\
    \lvert y\rvert<&~1.5\times10^{-5}~.
\end{align}
Here we focus on isocurvature perturbations and explore possible constraints on the parameter space spanned by the prior listed in Table~\ref{tab:priors}. Near-scale invariant adiabatic perturbations are known to be well consistent with FIRAS measurements and therefore we do not consider them here, see e.g.~\cite{Schoneberg:2020nyg,Cyr:2023pgw,Tagliazucchi:2023dai}.  We calculate the $\mu$- and $y$-type distortions of the CMB frequency spectrum as:
\begin{equation}
    \label{eq:sd}
    \mu,\,y=  \int_{k_{\rm min}}^\infty  {\rm d}\ln k\,\mathcal{P}_\mathcal{II}(k) \, W^{\mu,\,y}(k)~.
\end{equation} 
The window functions $W^{\mu,\,y}$ should in principle be computed numerically, given some SD visibility functions. However, Ref.~\cite{Chluba:2013dna} provided accurate analytical fits to the numerical results\footnote{For the adiabatic case, these equations were shown to match the exact numerical result obtained using the Green's function formalism  very well, even beyond power-law primordial power spectra~\cite{Tagliazucchi:2023dai}.} that we employ here, so that the calculation of the SDs simply amounts to a 1-dimensional integral over the comoving momenta $k$. The window functions are given by~\cite{Chluba:2013dna}
\begin{align}
    W^{\mu}(k)&\approx  2.8 \,C^2(k) \left[
    \exp\left(-\frac{\left[\frac{\hat{k}}{1360}\right]^2}{1+\left[\frac{\hat{k}}{260}\right]^{0.3}+\frac{\hat{k}}{340}}\right) 
    - \exp\left(-\left[\frac{\hat{k}}{32}\right]^2\right)
    \right],
    \\
    W^{y}(k)&
    \approx  \frac{C^2(k)}{2}\,  \exp\left(-\left[\frac{\hat{k}}{32}\right]^2\right),
\end{align}
with $\hat{k}=k/k_{\rm min}=k /(1~\Mpc^{-1})$, and
where 
\begin{equation}
    \label{eq:C_SD}
    C^2(k)=\frac{3}{8} \left(\frac{\Omega_c}{\Omega_m}\right)^2\beta_\nu(k_{\rm eq}/k)^2.
\end{equation}
Eq.~\eqref{eq:C_SD} depends only weakly on the cosmological parameters, so in this appendix, to get an order-of-magnitude estimation of the spectral distortions generated in our scenario, we take the benchmark values $\Omega_m=0.26,\,\Omega_c=0.216,\,\beta_\nu=[1 - 4\Omega_\nu/(\Omega_\nu+\Omega_\gamma) /15]/2 =0.79$ and $k_{\rm eq}=10^{-2}\,\Mpc^{-1}$. 

\begin{figure}[ht]
    \begin{center}
    \includegraphics[width=.495\columnwidth]{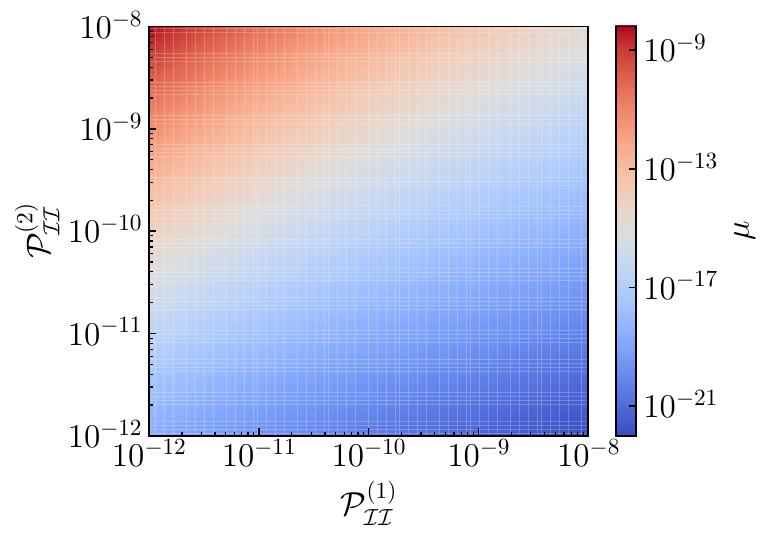}
    \includegraphics[width=.495\columnwidth]{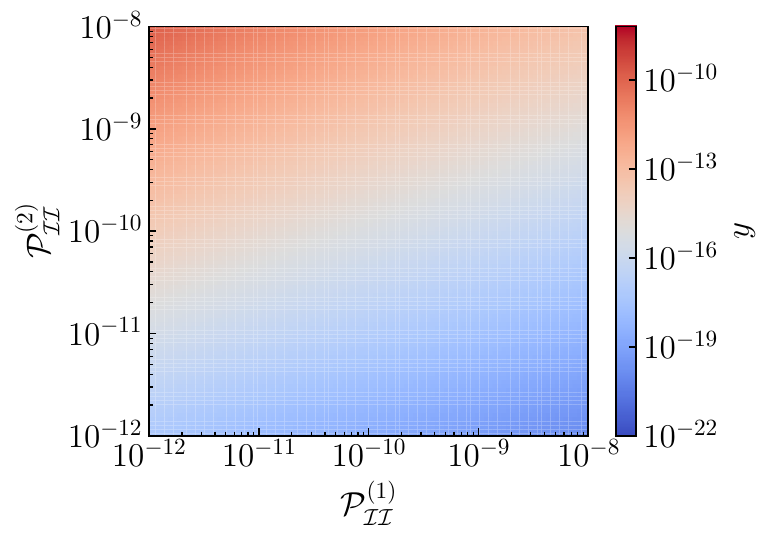}
    \end{center}
    \caption{\label{fig:sd_predictions} [Left] $\mu$-type and [right] $y$-type distortions for CDI perturbations. As can be seen, the parameter space spanned by the prior in Table~\ref{tab:priors} is well consistent with the limits $\lvert\mu\rvert<9.5\times10^{-5}$ and $\lvert y\rvert<1.5\times10^{-5}$ from FIRAS.}
\end{figure}

We plot the $\mu$- and $y$-type distortions for the parameter space explored in this paper (see Table~\ref{tab:priors}) in Figure~\ref{fig:sd_predictions}. 
As can be easily seen, all the points in our parameter space are well within the bounds from the FIRAS satellite.

\section{Full Constraints on Cosmological Parameters for CDI Models} \label{app:full_constraints}

Table~\ref{tab:cosmo_params} lists the 68\% C.L. constraints on cosmological parameters for each of the CDI scenarios studied in this paper: $\niso=0$, 1, 2, 3, and free $\niso$. The results for each of these models demonstrate consistency with the $\Lambda$CDM model and the constraints found by Refs.~\cite{Planck:2018vyg,ACT:2025fju,SPT-3G:2025bzu}. Note that the central values of the constraints on $\omega_b$, $\logA$, and $n_{\rm s}$ shift with each dataset, with Planck resulting in the smallest values, P-ACT resulting in the largest values, and the P-SPT results lying between the other two datasets' values. We discuss correlations between the cosmological parameters, particularly the curvature perturbation spectral index $n_{\rm s}$, and the isocurvature parameters in Sec.~\ref{sec:results}.

\begin{table}[ht]
    \centering
    \tiny
    
    \begin{tabular}{|l|llllll|}
        \hline Model \& Data &  $\omega_b$ & $\omega_c$ & $h$ & $\taureio$ & $\ln (10^{10}A_{\rm s})$ & $n_{\rm s}$ \\
        \hline \multicolumn{7}{|l|}{ $\niso=0$} \\
        \hline 
        Planck & $0.02218\pm 0.00014        $ & $0.1195\pm 0.0011          $ & $0.6732\pm 0.0051          $ & $0.0562^{+0.0056}_{-0.0062}$& $3.044\pm 0.013            $ & $0.9639\pm 0.0039$ \\
        P-ACT & $0.022586\pm 0.000096$ & $0.1189\pm 0.0010$ & $0.6787\pm 0.0043$ & $0.0684^{+0.0066}_{-0.0079}$ & $3.074^{+0.013}_{-0.015}$ & $0.9737\pm 0.0030$  \\
        P-SPT & $0.02226\pm 0.00012$ & $0.1195\pm 0.0012$ & $0.6734\pm 0.0050$ & $0.0605^{+0.0058}_{-0.0073}$ & $3.057^{+0.012}_{-0.014}$ & $0.9691\pm 0.0034$ \\
        \hline \multicolumn{7}{|l|}{ $\niso=1$} \\
        \hline 
        Planck & $0.02219\pm 0.00015$ & $0.1194\pm 0.0012$ & $0.6737\pm 0.0054$ & $0.0575^{+0.0054}_{-0.0065}$ & $3.046\pm 0.013$ & $0.9645\pm 0.0042$  \\
        P-ACT & $0.022583\pm 0.000097$ & $0.1188\pm 0.0011$ & $0.6791\pm 0.0045$ & $0.0691^{+0.0066}_{-0.0081}$ & $3.075^{+0.013}_{-0.015}$ & $0.9744\pm 0.0031$ \\
        P-SPT & $0.02225\pm 0.00011$ & $0.1195\pm 0.0011$ & $0.6735\pm 0.0047$ & $0.0610^{+0.0055}_{-0.0068}$ & $3.058^{+0.012}_{-0.014}$ & $0.9695\pm 0.0033$ \\
        \hline \multicolumn{7}{|l|}{ $\niso=2$} \\
        \hline 
        Planck & $0.02220\pm 0.00014        $  & $0.1193\pm 0.0012          $ & $0.6741\pm 0.0055          $ & $0.0575^{+0.0054}_{-0.0064}$& $3.045^{+0.012}_{-0.014}   $ & $0.9646\pm 0.0044          $ 
        \\
        P-ACT & $0.022594\pm 0.000097$ & $0.1186\pm 0.0010$ & $0.6799\pm 0.0044$ & $0.0695\pm 0.0072$ & $3.075\pm 0.014$ & $0.9750\pm 0.0032$  \\
        P-SPT & $0.02227\pm 0.00012$ & $0.1194\pm 0.0011$ & $0.6740\pm 0.0049$ & $0.0612^{+0.0058}_{-0.0068}$ & $3.058^{+0.011}_{-0.014}$ & $0.9700\pm 0.0034$ \\
        \hline \multicolumn{7}{|l|}{ $\niso=3$} \\
        \hline 
        Planck & $0.02222\pm 0.00015$ & $0.1191\pm 0.0013$ & $0.6749\pm 0.0056$ & $0.0571^{+0.0054}_{-0.0064}$ & $3.043^{+0.012}_{-0.014}$ & $0.9644\pm 0.0042$ \\
         P-ACT & $0.02261\pm 0.00010$ & $0.1186\pm 0.0011$ & $0.6803\pm 0.0046$ & $0.0688^{+0.0068}_{-0.0078}$ & $3.072\pm 0.014$ & $0.9742\pm 0.0031$  \\
         P-SPT & $0.02229\pm 0.00011$ & $0.1192\pm 0.0012$ & $0.6752\pm 0.0050$ & $0.0618^{+0.0055}_{-0.0068}$ & $3.057^{+0.012}_{-0.014}$ & $0.9696\pm 0.0035$ \\
         
        \hline \multicolumn{7}{|l|}{$\niso$ free} \\
        \hline
        Planck & $0.02221\pm 0.00015$ & $0.1191\pm 0.0012$ & $0.6748\pm 0.0055$ & $0.0572^{+0.0051}_{-0.0065}$ & $3.043^{+0.012}_{-0.014}$ & $0.9649\pm 0.0043$ \\
         P-ACT & $0.022592\pm 0.000098$ & $0.1186\pm 0.0011$ & $0.6801\pm 0.0045$ & $0.0693^{+0.0063}_{-0.0079}$ & $3.074^{+0.013}_{-0.015}$ & $0.9751\pm 0.0031$  \\
         P-SPT & $0.02228\pm 0.00012        $ & $0.1192\pm 0.0011          $ &$0.6749\pm 0.0048          $  & $0.0611^{+0.0054}_{-0.0066}$ & $3.056^{+0.012}_{-0.013}   $ & $0.9705\pm 0.0034          $ \\
        \hline
    \end{tabular}
    \caption{\normalsize 68\% C.L. bounds on cosmological parameters for the CDI models from Planck,~P-ACT,~and P-SPT datasets.} 
    \label{tab:cosmo_params}
\end{table}

\bibliography{Ref}

\end{document}